# Systematic external evaluation of published population pharmacokinetic models for tacrolimus in adult liver transplant recipients

Xiaojun Cai (1), Ruidong Li (2), Changcheng Sheng (1,3), Yifeng Tao (2), Quanbao Zhang (2), Xiaofei Zhang (2), Juan Li (2), Conghuan Shen (2), Xiaoyan Qiu (1), Zhengxin Wang (2) *, Zheng Jiao (1)*

((1) Department of Pharmacy, Huashan Hospital, Fudan University, Shanghai, China; (2) Liver Transplant Centre, Department of General Surgery, Huashan Hospital, Fudan University, Shanghai, China;(3) Department of Pharmacy, Guizhou Provincial People's Hospital, Guiyang, China)

**\*Corresponding author**

Professor Zhengxin Wang, M.D

Liver Transplant Centre, Department of General Surgery, Huashan Hospital, Fudan University, Shanghai, China

Email: wangzhengxin@huashan.org.cn

Professor Zheng Jiao, Ph.D

Department of Pharmacy, Huashan Hospital, Fudan University, Shanghai, China;

Email: zjiao@fudan.edu.cn

**Abstract**

*Background*

Diverse tacrolimus population pharmacokinetic (popPK) models in adult liver transplant recipients have been established to describe the PK characteristics of tacrolimus in the last two decades. However, their extrapolated predictive performance remains unclear. Therefore, in this study, we aimed to evaluate their external predictability and identify their potential influencing factors.

*Methods*

The external predictability of each selected popPK model was evaluated using an independent dataset of 84 patients with 572 trough concentrations prospectively collected from Huashan Hospital. Prediction- and simulation-based diagnostics and Bayesian forecasting were conducted to evaluate model predictability. Furthermore, the effect of model structure on the predictive performance was investigated.

*Results*

Sixteen published popPK models were assessed. In prediction-based diagnostics, the prediction error within ± 30% was below 50% in all the published models. The simulation-based normalised prediction distribution error test and prediction- and variability-corrected visual predictive check indicated large discrepancies between the observations and simulations in most of the models. Bayesian forecasting showed improvement in model predictability with two to three prior observations. Additionally, the predictive performance of the nonlinear Michaelis–Menten model was superior to that of linear one- and two-compartment models with first-order elimination, indicating the underlying nonlinear kinetics of tacrolimus in liver transplant recipients, which was consistent with the findings in adult kidney transplant recipients.

*Conclusions*

The published models performed inadequately in prediction- and simulation-based diagnostics. Bayesian forecasting may improve the predictive performance of the models. Furthermore, nonlinear kinetics of tacrolimus may be mainly caused by the properties of the drug itself, and incorporating nonlinear kinetics may be considered to improve model predictability.

*Keywords*

tacrolimus, population pharmacokinetics, adult liver transplant recipient, external evaluation, nonlinear kinetics

# 1 Introduction

Tacrolimus, a potent immunosuppressant of the calcineurin inhibitor family, is considered the cornerstone of the prevention of graft rejection in solid organ-transplanted patients (Siekierka et al., 1989). Currently, it is the drug of primary choice in almost 90% of liver transplantation patients and tends to replace cyclosporine A because of its better long-term graft and patient survival rates (European Association for the Study of the Liver. Electronic address, 2016).

Following oral administration, tacrolimus is rapidly absorbed and reaches the peak concentration 0.5–1 h later with a mean bioavailability of 25% (5%–93%) in patients (Staatz and Tett, 2004; Vanhove et al., 2016b). After entering the systemic circulation, tacrolimus is extensively bound to erythrocytes and albumin (about 99%) (Yu et al., 2018). Tacrolimus is primarily metabolised by cytochrome P450 (CYP) 3A isoenzymes in the liver and intestine, and transported by the efflux pump P-glycoprotein (P-gp) encoded by the ATP-binding cassette, sub-family B member 1 (*ABCB1*) gene (Vanhove et al., 2016b). In addition, it is excreted predominantly through bile (> 95%), with renal clearance accounting for only 2.4% of elimination (European Association for the Study of the Liver. Electronic address, 2016; Staatz and Tett, 2004).

Tacrolimus is characterised by a narrow therapeutic window and exhibits considerable intra- and inter-individual variabilities in pharmacokinetics, which can be attributed to multiple factors, including postoperative time, polymorphisms in drug-metabolizing enzymes and transporters such as *CYP3A5* and *ABCB1*, patient demographics, daily dose, graft type (whole or split liver), hepatic function, international normalised ratio (INR), graft to recipient body weight ratio, and concomitant medications (Campagne et al., 2019; Ji et al., 2012). Moreover, tacrolimus trough levels are closely associated with the risk of acute liver allograft rejection and adverse effects, such as infection, nephrotoxicity, hypertension, post-transplant diabetes mellitus, and gastrointestinal disorders (Vanhove et al., 2016a), necessitating therapeutic drug monitoring (TDM), especially during the early post-transplant period (Shuker et al., 2015).

Monitoring of tacrolimus, as a traditional approach, is widely performed to ensure the maintenance of a functional allograft and to minimise toxicity, but it has practical difficulties, such as the impossibility to explain individual variation in transplant patients (Zhang et al., 2019). Population pharmacokinetics (popPK) is a superior approach to conventional PK analysis, and has become a potent and more reliable tool used to facilitate the quantification and explanation of PK variability and to identify the sources of variability (Brooks et al., 2016; Mao et al., 2018; Vadcharavivad et al., 2016). Combined with Bayesian estimations, it can guide tacrolimus dose adaption more precisely and rapidly to reach the target concentration than that based on the personal experience of transplant physicians alone (Mizuno et al., 2019). Therefore, it plays an important role in optimising tacrolimus dosage regimen.

Numerous tacrolimus popPK models have been established in adult liver transplant recipients for quantitatively describing the PK characteristics and dosing individualisation (Campagne et al., 2019). However, their predictability remains unknown when extrapolated to other clinical centres. Consequently, it is necessary to systematically investigate the cross-centre predictability of these published models with an external dataset. Additionally, evaluating the model transferability may contribute to the identification of potential factors responsible for model predictability and reveal whether previous popPK studies can be directly used to guide tacrolimus dosing recommendations. Moreover, nonlinear kinetics of tacrolimus was observed in adult renal transplant recipients (Zhao et al., 2016). Whether the nonlinear kinetics is also valid in liver transplantation remains unknown, and thus, further analysis needs to be conducted to confirm this.

Therefore, the aim of the present study was to assess the external predictability of published popPK models of tacrolimus in adult patients receiving liver transplantation using an independent dataset prospectively collected at our centre. Additionally, potential influencing factors, which are mainly responsible for the specification of model predictability, were investigated.

## 2 Methods

### 2.1 Review of published popPK analysis on tacrolimus

To identify publications of all popPK analyses of tacrolimus in adult liver transplant recipients, PubMed, Web of Science, and Embase were searched for human data in English from inception to June 30, 2019. Furthermore, additional articles identified from the reference lists of selected publications were also screened. The eligibility criteria for publications were as follows: (1) studies involving adult patients who underwent liver transplantation and treated with oral tacrolimus, (2) articles involving popPK modelling restricted to population analyses. Reviews, administration of tacrolimus in a prolonged-release formulation, methodological articles, incomplete or missing information of required details for external evaluation, and unavailable covariate information in our evaluation dataset were excluded. Studies with overlapping data or cohorts were also excluded. Only the most recent studies or those with a large sample size were included.

### 2.2 Dataset of external evaluation

#### 2.2.1 Subjects

Eighty-four adult recipients (69 males/15 females) who underwent their first liver transplantation using organs from donation after cardiac death and were administered immediate-release oral tacrolimus formulation (Prograf, Astellas, Dublin, Ireland) in Huashan Hospital of Fudan University from June 2018 to January 2019 were included in this study. Recipient follow-up was continued until the day of discharge. Only monitored records after more than three repeated oral doses administered at the same dose rate were retrieved to ensure that tacrolimus concentrations were at or near the steady state. Finally, 572 tacrolimus whole-blood trough concentrations ($C_0$) were prospectively collected during the eligible patients' hospitalisation. Patients with acute rejection, dialysis treatment, or severe gastrointestinal disorders, or those who underwent their secondary liver transplantation were excluded. The study protocol was approved by the Ethics Committee of Huashan Hospital, Fudan University, and registered at the Centre for Clinical Research and Biostatistics (www2.ccrb.edu.hk, No: CUHK_TMP00250). Informed written consent was obtained from all subjects before enrolment. Additionally, the study was conducted in accordance with the Declaration of Helsinki (2013).

*2.2.2 Immunosuppressive therapy*

All patients received immunosuppressive therapy comprising tacrolimus and steroids after liver transplantation. Oral tacrolimus administration was commenced at 0.5–1 mg every 12 h (q12h). The dosage was then empirically adjusted to attain steady-state $C_0$ within the range of 8–12 ng ml$^{-1}$ in the first 3 months post-operation, 8–10 ng ml$^{-1}$ between 3 and 6 months, and 6–8 ng ml$^{-1}$ thereafter.

Intravenous methylprednisolone was administered at a dose of 500 mg on the operative day, followed by 80 mg q12h on postoperative days 1–3. Then, the dosage was tapered to 80 mg day$^{-1}$ on postoperative days 4 and 5 followed by 40 mg day$^{-1}$ on postoperative days 6 and 7. Thereafter, the dosage was reduced to 20 mg day$^{-1}$ on postoperative days 8–10. Oral prednisolone was administered at a dosage of 12 mg day$^{-1}$ on postoperative day 11 and was tapered to 4 mg day$^{-1}$ at a rate of 4 mg day$^{-1}$, except for hepatocellular carcinoma patients with liver transplantation. During the second month after surgery, corticosteroid-free treatment was conducted, except for patients with autoimmune hepatitis.

Mycophenolate mofetil (CellCept, Roche Pharma Ltd., Shanghai, China) was administered orally q12h at 0.5 g day$^{-1}$ to patients who had glomerular filtration rate (GFR) of less than 60 ml min$^{-1}$ per 1.73 m$^2$ (Hao et al., 2014).

*2.2.3 Blood sample collection and bioassay*

After reaching a steady-state condition, whole blood samples were drawn before the morning dose for assaying $C_0$ using the enzyme multiplied immunoassay technique (EMIT) with SYVA Viva-Emit 2000 kit (Siemens Healthcare Diagnostics Inc., Germany). The coefficient of variation (% CV) of intra- and inter-day precisions was within 10% and 20%, respectively, with a calibration range between 2.0 and 30 ng ml$^{-1}$.

Diverse bioassays were performed in previous studies and systematic deviations existed among the different analysis methods (Agrawal et al., 2014; Bazin et al., 2010; LeGatt et al., 2004). To adjust these biases, tacrolimus concentrations in the external dataset were converted to their corresponding equivalents with the following formulas from large clinical studies in adult liver transplant recipients:

$$LCMS = \frac{EMIT - 0.01}{1.37} \qquad (1) \qquad \text{(Ansermot et al., 2008)}$$

$$MEIA = \frac{EMIT - 0.25}{1.00} \qquad (2) \qquad \text{(Hesse et al., 2002)}$$

Where *EMIT* is the concentration of the external dataset determined using the EMIT, and *LCMS* and *MEIA* are the after-conversion equivalents analysed by liquid chromatography-mass spectrometry (LC-MS) and microparticle enzyme immunoassay (MEIA), respectively.

*2.2.4 Genotyping*

Whole blood (2 ml) was obtained from liver transplant recipients and their corresponding donors, and then drawn into sterile ethylene dinitrilotetraacetic acid (EDTA)-anticoagulated tubes, which were stored at -20 °C. Genetic polymorphisms of *CYP3A5\*3* (*rs776746*) was determined by an independent external contractor (Sangon Biotechnology Co., Ltd., Shanghai, China) using the DNA direct sequencing Analyzer (Applied Biosystems 3730XL, Foster City, CA, USA). Alleles and genotype frequencies were analysed using the online software SHEsis (http://analysis.bio-x.cn/myAnalysis.php). The Hardy–Weinberg equilibrium of genetic polymorphisms was assessed using Pearson's chi-squared test. Details of gene amplification and sequencing are presented in Appendix S1.

*2.3 External evaluation of predictability*

The external evaluation was performed using nonlinear mixed-effects modelling software (NONMEM®, version 7.4; ICON Development Solutions, Ellicott City, MD, USA) compiled with gfortran 4.6.0 and interfaced with PsN (version 4.7.0; uupharmacometrics.github.io/PsN). The NONMEM output was analysed using the R software (version 3.5.1; www.r-project.org). The reported popPK models were re-established based on the formulas and parameters extracted from each identified article. The codes for the models were sent to the corresponding authors for crosscheck. Subsequently, the external evaluation of predictability for all the candidate models was conducted with prediction- and simulation-based diagnostics, and Bayesian forecasting.

*2.3.1 Prediction-based diagnostics*

Population predictions (PREDs) were estimated and compared to the corresponding observations (OBS) based on the prediction error (PE%) calculated using equation (3).

$$PE\% = \left(\frac{PRED - OBS}{OBS}\right) \times 100 \qquad (3)$$

The median prediction error (MDPE) and median absolute prediction error (MAPE) were applied to investigate the accuracy and precision of predictability, respectively. $F_{20}$ (PE% within ± 20%) and $F_{30}$ (PE% within ± 30%), an index of both accuracy and precision, were also calculated. If a model reached the criteria of MDPE ≤ ± 20%, MAPE ≤ 30%, $F_{20}$ ≥ 35%, and $F_{30}$ ≥ 50% (Mao et al., 2018; Zhang et al., 2019), its predictive performance was considered to be satisfactory and clinically acceptable.

*2.3.2 Simulation-based diagnostics*

To assess the predictability of each selected model based on simulation, prediction- and variability-corrected visual predictive check (pvcVPC) as well as normalised prediction distribution error (NPDE) test were conducted to compare the simulated and observed data. The dataset was simulated 2000 times using the $SIMULATION module of NONMEM®.

The calculations and graphical visualisations for pvcVPC were performed with PsN. To identify systematic bias between the observed and simulated data, the 95% confidence intervals (CIs) for the median and the 5th and 95th percentiles of the simulated concentrations at different bins were calculated and compared with the observations. The bins were automatically selected by the VPC command in PsN. NPDE determination was implemented with the NPDE add-on package in R (version 2.0; www.npde.biostat.fr). Based on the null hypothesis that the evaluation data can be well described by a candidate model, NPDE follows a standard normal distribution.

*2.3.3 Bayesian forecasting*

Maximum *a posteriori* Bayesian (MAPB) forecasting was conducted using data from patients with ≥ 5 observations to evaluate the effect of priors on model predictive performance. For each patient, the individual prediction (IPRED) of the fifth observation was predicted by the last one, two, three, and four prior observations, and subsequently compared with the corresponding observation. Individual prediction error (IPE%) was calculated using equation (4) as follows:

$$IPE\% = (\frac{IPRED - OBS}{OBS}) \times 100 \qquad (4)$$

To assess the predictability of a candidate model as the prior information increased, median IPE% (MDIPE), median absolute IPE% (MAIPE), and $IF_{20}$ and $IF_{30}$, that represented $F_{20}$ and $F_{30}$ of IPE%, respectively, were computed under the circumstance of nought to four priors.

*2.4 Influence of model structures*

Considering that model structure was one of the primary factors affecting the predictive performance, the various structural models reported in previous studies were reviewed (Zhao et al., 2016). The prediction-based diagnostics and Bayesian forecasting described above were employed to investigate the predictability of structural models.

3 Results

*3.1 Review of published popPK analysis on tacrolimus*

Sixteen popPK models for tacrolimus in adult liver transplant recipients (Antignac et al., 2005; Blanchet et al., 2008; Chen et al., 2017; Dansirikul et al., 2004; Fukudo et al., 2003; Ji et al., 2018; Lee et al., 2006; Li et al., 2007; Lu et al., 2015; Oteo et al., 2013; Sam et al., 2006; Staatz et al., 2003; Zahir et al., 2005; Zhang et al., 2012; Zhu et al., 2014; Zhu et al., 2015) were finally included for external evaluation after the literature search procedure. The details are presented in Appendix S2. Information in all the identified studies are summarised in Table 1 and Appendix S3.

Among the identified studies, 10 of them established a popPK model only with trough concentrations (Antignac et al., 2005; Fukudo et al., 2003; Ji et al., 2018; Lee et al., 2006; Li et al., 2007; Lu et al., 2015; Oteo et al., 2013; Zahir et al., 2005; Zhang et al., 2012; Zhu et al., 2015). In the remaining studies, four of them used an intensive sampling strategy with ≥ 6 samplings (Blanchet et al., 2008; Chen et al., 2017; Sam et al., 2006; Zhu et al., 2014), whereas the others adopted a limited strategy using $C_0$ and additional 2–3 samplings (Dansirikul et al., 2004; Staatz et al., 2003).

The popPK model structures used in all the candidate analyses were linear one-compartment (1CMT) (Antignac et al., 2005; Fukudo et al., 2003; Ji et al., 2018; Lee et al., 2006; Li et al., 2007; Oteo et al., 2013; Sam et al., 2006; Staatz et al., 2003; Zahir et al., 2005; Zhang et al., 2012; Zhu et al., 2015) or two-compartment (2CMT) (Blanchet et al., 2008; Chen et al., 2017; Dansirikul et al., 2004; Lu et al., 2015; Zhu et al., 2014) models with first-order absorption and elimination. In 56.2% of these studies (Antignac et al., 2005; Ji et al., 2018; Li et al., 2007; Lu et al., 2015; Oteo et al., 2013; Staatz et al., 2003; Zahir et al., 2005; Zhang et al., 2012; Zhu et al., 2015), the absorption rate constant (Ka) was fixed, as only tacrolimus trough concentrations were collected.

Several covariates were found to have a significant effect on the pharmacokinetics of tacrolimus. Among them, postoperative days (POD), haematocrit (HCT), and total bilirubin (TBIL) were the most frequently identified covariates influencing apparent clearance (CL/F), as reported in eight (Antignac et al., 2005; Chen et al., 2017; Fukudo et al., 2003; Ji et al., 2018; Lee et al., 2006; Oteo et al., 2013; Zhu et al., 2014; Zhu et al., 2015), four (Oteo et al., 2013; Zahir et al., 2005; Zhang et al., 2012; Zhu et al., 2015) and four studies (Fukudo et al., 2003; Lee et al., 2006; Li et al., 2007; Zhu et al., 2015), respectively. Moreover, POD and body weight (BW) were the most frequent covariates influencing apparent volume of distribution (V/F), incorporated in three (Ji et al., 2018; Oteo et al., 2013; Zhu et al., 2015) and two studies (Fukudo et al., 2003; Staatz et al., 2003), respectively. Other covariates identified in the published models were tacrolimus daily dose, age, height (HT), haemoglobin (HB), aspartate aminotransferase (AST), alanine aminotransferase (ALT), total protein (TP), albumin (ALB), serum creatinine (SCR), creatinine clearance (CCR), *CYP3A5* genotype of both recipients and donors, genetic polymorphisms of *ABCB1 (C3435T)* of recipients, concomitant diltiazem, fluconazole or sulfonylurea co-administration, and graft to recipient body weight ratio (GRWR).

## 3.2 External evaluation cohort

A total of 572 trough concentrations from 84 patients from 4 to 50 days after transplantation were included in this study. All the demographic and laboratory test data, concomitant medications, and primary diseases collected for the evaluation are summarised in Table 2. Thirteen observations were below the lower limit of quantification (2.0 ng ml$^{-1}$) and they were included in the analysis as the original reported values. In our dataset, one patient was co-administered diltiazem, five patients were co-administered mycophenolate mofetil, and seven patients were co-administered fluconazole. No patient was co-administered sulfonylureas. In addition, the allele frequencies of *CYP3A5*3* genetic polymorphisms are listed in Table 3 and were in Hardy–Weinberg equilibrium. As the *ABCB1* genotype was not collected in our data, the genotype of *ABCB1 C3435T* was assumed to be *ABCB1 3435CC*, which has the highest frequency in the Chinese population.

## 3.3 External predictability evaluation

### 3.3.1 Prediction-based diagnostics

The results of prediction-based diagnostics are provided in Figure 1 and Table S1. None of the investigated models met all the aforementioned standards (MDPE ≤ ± 20%, MAPE ≤ 30%, $F_{20}$ ≥ 35%, and $F_{30}$ ≥ 50%), indicating unsatisfactory predictive performance. MDPE, as an index of predictive accuracy, was less than ± 20% in six studies (Ji et al., 2018; Lee et al., 2006; Li et al., 2007; Lu et al., 2015; Oteo et al., 2013; Zhang et al., 2012), whereas MAPE, as an index of predictive precision, was more than 30% in all studies. Considering both the accuracy and precision of predictability, the model reported by Zhang et al.(Zhang et al., 2012) was superior to the others, with $F_{20} > 35\%$ and $F_{30} > 45\%$.

*3.3.2 Simulation-based diagnostics*

In simulation-based diagnostics, pvcVPC showed a large discrepancy between the observations and model simulations in all published studies (Figure S2). A significant trend of over- or under-prediction was observed, indicating misspecifications of the models. The relatively superior model by Zhang in the prediction-based diagnosis (Zhang et al., 2012) also performed poorly.

The NPDE results are presented in Figure S3 and Table S2. NPDE distribution of all studies significantly deviated from the standard normal distribution. All models were rejected with no adjusted *P* values over 0.01 for the global test.

*3.3.3 Bayesian forecasting*

The results of Bayesian forecasting proved that prior observations significantly improved both predictive precision and accuracy even with only one prior observation. Predictive performance reached a stable state with two or three priors. More number of priors did not achieve further obvious improvement. Furthermore, in this evaluation, the models of Zhang et al.(Zhang et al., 2012), Zhu et al.(Zhu et al., 2014), and Zhu et al.(Zhu et al., 2015) were the best three, which showed MDIPE < 20%, MAIPE < 30%, $IF_{20} > 35\%$ and $IF_{30} > 50\%$ under the circumstance of two to four priors. The box plots of predictability are presented in Figure 2, and the results of IPE% are listed in Table S4.

*3.4 Effect of model structures*

The model structure employed in all the included studies was either linear 1CMT or 2CMT model. As nonlinear kinetics of tacrolimus was observed in kidney transplant recipients (Zhao et al., 2016), three covariate-free model structures, namely the linear 1CMT, linear 2CMT, and nonlinear Michaelis–Menten (MM) model, were evaluated. The nonlinear MM empirical model was used to quantify the relationship between daily dose and $C_0$, presented as follows (Zhao et al., 2016):

$$Daily\ dose\ (mg\ day^{-1}) = \frac{V_m(mg) \times C_0(ng\ mL^{-1})}{K_m(ng\ mL^{-1}) + C_0(ng\ mL^{-1})} \quad (5)$$

Where $V_m$ denotes the maximum dose rate (daily dose) at the steady state. The Michaelis constant, $K_m$, denotes the steady-state trough concentration at half-maximal dose rate. The estimated parameters for all the three structural models are presented in Table S3.

The results of prediction-based diagnostics shown in Figure 3 and Table S1 indicated that the nonlinear MM model performed better than the linear 1CMT and 2CMT models. The $F_{20}$ and $F_{30}$ of the covariate-free MM model showed an improvement over those of the linear 1CMT (29.72% vs. 25.87% and 25.87%) and 2CMT models (43.01% vs. 38.99% and

40.21%).

The result of Bayesian forecasting listed in Figure 4 and Table S4 showed that the predictive performance of each model was improved with MAPB even with only one prior observation. The predictability reached a stable state with two or three priors and improvement cannot be further achieved with more prior observations. The $IF_{20}$ and $IF_{30}$ values of the covariate-free MM model after Bayesian forecasting with one prior reached 49% and 74%, respectively, which were considerably better than those of the linear 1CMT model (31% and 46%, respectively) and 2CMT model (29% and 48%, respectively). Furthermore, with two to four priors, the $IF_{20}$ and $IF_{30}$ values of the covariate-free MM model after Bayesian forecasting were all beyond 35% and 50%, respectively, whereas both linear models failed to attain the criteria of $IF_{20} \geq 35\%$ and $F_{30} \geq 50\%$.

## 4 Discussion

To the best of our knowledge, this is the first comprehensive external evaluation of published tacrolimus popPK models in adult liver transplant patients using an independent dataset, which was prospectively collected from routine TDM, ensuring the accuracy of the data record.

The results showed that the predictive accuracy was acceptable with MDPE $\leq \pm 20\%$ in six studies. However, poor predictive precision was observed, and the MAPE fell outside $\pm 30\%$ in all the investigated models. Therefore, taking both accuracy and precision into account, the prediction-based predictive performances were unsatisfactory. In addition, the simulation-based pvcVPC and NPDE tests showed that all the published models failed to fulfil the diagnostic criteria.

Considering that the predictive performance depended largely on the model structure and structural model components had a large effect on the clinical utility of model-based personalised dosing (Mao et al., 2018; McDougall et al., 2016; Zhang et al., 2019), the influence of model structures on predictability was investigated. The predictive performance of the nonlinear MM model was shown to be superior to those of linear 1CMT and 2CMT models, which supported the nonlinearity of tacrolimus PK in adult liver transplant recipients. This finding is consistent with our previous report on tacrolimus PK in adult renal transplant patients (Zhao et al., 2016).

The identified nonlinear kinetics of tacrolimus PK may be partly attributed to its poor aqueous solubility (1–2 μg ml$^{-1}$) (Lee et al., 2016) and low permeability to the intestinal membrane (Tamura et al., 2002). These lead to a dissolution rate-limited absorption in the gastrointestinal tract, and variable and low oral bioavailability (Lee et al., 2016). Additionally, the recovery of gastrointestinal function, metabolising enzymes, and P-gp activity with POD, as well as the gradually decreased induction of CYP3A enzymes and P-gp in the liver and intestine by tapering of co-administered steroid doses, may also lead to the nonlinear pharmacokinetic behaviour (Abuasal et al., 2012; Christians et al., 2002; Shimada et al., 2002; Tubic et al., 2006).

Moreover, a blockage of albumin synthesis caused by poor hepatic function and a low HCT level in the early postoperative period may lead to saturated tacrolimus concentration-dependent binding to albumin and erythrocytes (Chow et al., 1997). Biliary complications, such as biliary strictures and bile poor healing in liver transplant recipients, usually require external biliary drainage, which increases drug excretion via bile routine (Sarhan et al., 2017; Taner et al., 2012). These complications contributed to the nonlinearity in the distribution and elimination of tacrolimus.

However, the nonlinearity of tacrolimus in liver transplant patients may be different from that in renal transplant patients. In renal transplant patients, with 1–4 prior observations, the $IF_{20}$% and $IF_{30}$% of the covariate-free MM models after MAPB forecasting reached 65%–90% and 94%–100%, respectively, whereas those in liver transplant patients achieved 35%–49% and 57%–74%, respectively. This is most likely due to the poor hepatic function significantly affecting the first-pass metabolism of tacrolimus and different corticosteroid dosage used in patients with liver transplantation (Bekersky et al., 2001; Shimomura et al., 2002; Vanhove et al., 2016a). Additionally, most of the liver transplant recipients included in our study were hepatitis B-positive (n = 61), hepatitis C-positive (n = 3), or hepatitis E-positive (n = 1). Replication of hepatitis virus in hepatocytes alters the CYP3A system, leading to reduced tacrolimus metabolism (Horina et al., 1993), which is different to that in renal transplant recipients. Further investigations need to be conducted to explore the underlying mechanism.

Beyond those mentioned above, the TDM effect is also considered a factor leading to nonlinearity. Individuals with higher drug clearance usually have lower drug concentrations. This will make clinicians prescribe higher doses. Therefore, a relationship between dose and clearance will be finally induced during the TDM process, which may be misclassified as nonlinearity (Ahn et al., 2005; ES et al., 2004). Nevertheless, if the subjects are sampled at more than three dose levels, such as in the current external dataset, the TDM effect will not influence the identification of nonlinearity (Ahn et al., 2005).

In addition to the model structures discussed above, covariates may also be responsible for predictive performance (Vanhove et al., 2016a). POD and HCT were the two prevailing covariates identified to affect the CL/F of tacrolimus in liver transplant recipients.

POD was identified as a major surrogate for several time-dependent variables (Ette and Ludden, 1995). The gradual improvement in metabolic function with POD increased the CL/F during the early stage after liver transplantation (Chen et al., 2017). Therefore, an increase in the dose of tacrolimus was usually required to maintain similar trough concentrations with increasing time after transplantation (Wallin et al., 2011). In addition, tapering of corticosteroid doses with POD is known to decrease the CL/F of tacrolimus, leading to a decrease in the dose of tacrolimus to maintain similar trough concentrations with increasing time after transplantation (Passey et al., 2011). The effect of POD on tacrolimus PK may also be related to other factors. Their simultaneous inclusion as covariates in the final model would definitely affect the description of POD.

Tacrolimus mainly binds to erythrocytes, and changes in HCT alter the distribution of tacrolimus between blood and fat because it is a lipophilic drug (Sam et al., 2000). Low HCT values increase the partitioning of tacrolimus into fat, thereby leading to a higher apparent volume of distribution, which reflects a greater availability of unbound tacrolimus for distribution in peripheral tissues (Press et al., 2009; Sam et al., 2000; Sam et al., 2006). As the unbound tacrolimus increased, relatively more tacrolimus is metabolized and eliminated from the body, resulting in increased CL/F. The relationship between HCT and tacrolimus PK has been well described in published models, and a low HCT of < 35% results in 46% higher CL/F than HCT levels of > 35% (Zahir et al., 2005).

*CYP3A5\*3* genetic polymorphisms is also considered a predominant factor associated with tacrolimus PK (Goto et al., 2004), and its polymorphisms in both donors and recipients partially explain the between-subject variability in tacrolimus clearance (Li et al., 2007; Zhu et al., 2015). The CL/F in recipients with the *CYP3A5\*1* allele grafted from a donor with and without the *CYP3A5\*1* allele, respectively, was 2.3- and 1.5-fold higher than that in recipients with the *CYP3A5\*3/\*3* genotype (Ji et al., 2018). When the *CYP3A5\*3* genotypes of both recipients and donors were taken into consideration, the between-subject variability of CL/F was reduced from 40.6% to 31.2% (Li et al., 2007). Moreover, tacrolimus is transported by P-gp encoded by the *ABCB1* gene, but polymorphism of *ABCB1* was not considered as a covariate in most studies (Ji et al., 2018; Oteo et al., 2013; Staatz et al., 2003).

Additionally, predicting a dose without TDM in a patient will not be accurate (Tauzin et al., 2019). However, through Bayesian forecasting, dosing can be predicted and adjusted more precisely even with one prior sample (Mizuno et al., 2017; Sheiner and Beal, 1982). External evaluation of the performance of the published models with Bayesian forecasting may help find the most appropriate model and its related influencing factors, as well as guide dose adjustment in clinical practice. In this study, the results of prediction-based and simulation-based diagnostics showed the poor predictive performance of the published models. Bayesian forecasting showed that reliable and stable prediction was obtainable by MAPB estimation with 2–3 prior observations, which is in agreement with the results from renal transplant subjects (Mao et al., 2018; Zhang et al., 2019; Zhao et al., 2016). This indicated that population model-based approach using Bayesian forecasting in combination with TDM can be implemented to individually adjust tacrolimus dosing regimens.

This study had some limitations. One limitation is that we concentrated on the trough concentrations alone, which limited the evaluation of the absorption models, such as the transit models. However, based on the $C_0$, the CL/F of tacrolimus was still reliably estimated. Another limitation is the bioassay method, which is inconsistent with those used in most of the published studies. As the analytical method may result in bias of parameters estimation (Akbas et al., 2005; Laporte-Simitsidis et al., 2000), two conversional formulas were used to make the inter-method results equivalent. After conversion, all the models with different bioassay methods were grouped together, but no obvious bioassay-related trends were found in the predictive performance. This indicated that the difference in bioassay between the external dataset and the investigated model was not a dominant limitation after inter-method conversion. However, the inter-method biases may not be completely eliminated, which may have an effect on more or less predictive performance (Wallemacq et al., 2009; Zhao et al., 2016).

## 5. Conclusions

The external predictive performance of the 16 investigated population PK models for tacrolimus in adult recipients with liver transplantation was unsatisfactory in both prediction- and simulation-based diagnostics. The MAPB approach significantly improved both predictive accuracy and precision of the models, which can be used to guide tacrolimus dosing recommendations and adjustments for clinicians. Furthermore, this study highlights that the predictive performance of the MM model was superior to that of linear compartment models, indicating underlying nonlinear kinetics of tacrolimus in liver transplant patients. This may be attributed to the properties of tacrolimus itself. Further studies are required to confirm our findings.

**Acknowledgements**

The authors would like to sincerely thank *Dr. F. Akhlaghi* from the Department of Biomedical and Pharmaceutical Sciences, College of Pharmacy, University of Rhode Island (USA), and *Dr Bing Chen* from Ruijin Hospital of Shanghai Jiao Tong University School of Medicine (China), and *Dr Liqin Zhu* from the Department of Pharmacy, Tianjin First Central Hospital (China) for providing details about the research and active discussions on the coding. We would also like to thank Editage (www.editage.cn) for English language editing.

**Funding**

This work was supported by the National Natural Science Foundation of China (81573505), and the "Weak Discipline Construction Project" (2016ZB0301-01) of Shanghai Municipal Commission of Health and Family Planning.

**Declarations of interest**

None

**Author contributions**

Xiaojun Cai, Zheng Jiao, and Zhengxin Wang designed the research and planned the work that led to the manuscript. Xiaojun Cai, Ruidong Li, Yifeng Tao, Juan Li, Conghua Shen, and Xiaoyan Qiu acquired the data and were responsible for interpreting the data. Xiaojun Cai, Quanbao Zhang, and Xiaofei Zhang collected the blood samples. Xiaojun Cai and Changcheng Sheng analysed the data and provided statistical expertise. Xiaojun Cai and Zheng Jiao drafted the manuscript, which was commented on and approved by all the authors.

**References**

Abuasal, B.S., Bolger, M.B., Walker, D.K., Kaddoumi, A., 2012. In silico modeling for the nonlinear absorption kinetics of UK-343,664: a P-gp and CYP3A4 substrate. Mol Pharm 9, 492-504.

Agrawal, Y.P., Cid, M., Westgard, S., Parker, T.S., Jaikaran, R., Levine, D.M., 2014. Transplant patient classification and tacrolimus assays: more evidence of the need for assay standardization. Ther Drug Monit 36, 706-709.

Ahn, J.E., Birnbaum, A.K., Brundage, R.C., 2005. Inherent correlation between dose and clearance in therapeutic drug monitoring settings: possible misinterpretation in population pharmacokinetic analyses. J Pharmacokinet Pharmacodyn 32, 703-718.

Akbas, S.H., Ozdem, S., Caglar, S., Tuncer, M., Gurkan, A., Yucetin, L., Senol, Y., Demirbas, A., Gultekin, M., Ersoy, F.F., Akaydin, M., 2005. Effects of some hematological parameters on whole blood tacrolimus concentration measured by two immunoassay-based analytical methods. Clin Biochem 38, 552-557.

Ansermot, N., Fathi, M., Veuthey, J.L., Desmeules, J., Rudaz, S., Hochstrasser, D., 2008. Quantification of cyclosporine and tacrolimus in whole blood. Comparison of liquid chromatography-electrospray mass spectrometry with the enzyme multiplied immunoassay technique. Clin Biochem 41, 910-913.

Antignac, M., Hulot, J.S., Boleslawski, E., Hannoun, L., Touitou, Y., Farinotti, R., Lechat, P., Urien, S., 2005. Population pharmacokinetics of tacrolimus in full liver transplant patients: modelling of the post-operative clearance. Eur J Clin Pharmacol 61, 409-416.

Bazin, C., Guinedor, A., Barau, C., Gozalo, C., Grimbert, P., Duvoux, C., Furlan, V., Massias, L., Hulin, A., 2010. Evaluation of the Architect tacrolimus assay in kidney, liver, and heart transplant recipients. J Pharm Biomed Anal 53, 997-1002.

Bekersky, I., Dressler, D., Alak, A., Boswell, G.W., Mekki, Q.A., 2001. Comparative tacrolimus pharmacokinetics: normal versus mildly hepatically impaired subjects. J Clin Pharmacol 41, 628-635.

Blanchet, B., Duvoux, C., Costentin, C.E., Barrault, C., Ghaleh, B., Salvat, A., Jouault, H., Astier, A., Tod, M., Hulin, A., 2008. Pharmacokinetic-pharmacodynamic assessment of tacrolimus in liver-transplant recipients during the early post-transplantation period. Ther Drug Monit 30, 412-418.

Brooks, E., Tett, S.E., Isbel, N.M., Staatz, C.E., 2016. Population Pharmacokinetic Modelling and Bayesian Estimation of Tacrolimus Exposure: Is this Clinically Useful for Dosage Prediction Yet? Clin Pharmacokinet 55, 1295-1335.

Campagne, O., Mager, D.E., Tornatore, K.M., 2019. Population Pharmacokinetics of Tacrolimus in Transplant Recipients: What Did We Learn About Sources of Interindividual Variabilities? J Clin Pharmacol 59, 309-325.

Chen, B., Shi, H.Q., Liu, X.X., Zhang, W.X., Lu, J.Q., Xu, B.M., Chen, H., 2017. Population pharmacokinetics and Bayesian estimation of tacrolimus exposure in Chinese liver transplant patients. J Clin Pharm Ther 42, 679-688.

Chow, F.S., Piekoszewski, W., Jusko, W.J., 1997. Effect of hematocrit and albumin concentration on hepatic clearance of tacrolimus (FK506) during rabbit liver perfusion. Drug Metab Dispos 25, 610-616.

Christians, U., Jacobsen, W., Benet, L.Z., Lampen, A., 2002. Mechanisms of clinically relevant drug interactions associated with tacrolimus. Clin Pharmacokinet 41, 813-851.

Dansirikul, C., Staatz, C.E., Duffull, S.B., Taylor, P.J., Lynch, S.V., Tett, S.E., 2004. Sampling times for monitoring tacrolimus in stable adult liver transplant recipients. Ther Drug Monit 26, 593-599.

ES, E.L.D., Fuseau, E., S, E.L.D.A., Cosson, V., 2004. Pharmacokinetic modelling of valproic acid from routine clinical data in Egyptian epileptic patients. Eur J Clin Pharmacol 59, 783-790.

Ette, E.I., Ludden, T.M., 1995. Population pharmacokinetic modeling: the importance of informative graphics. Pharm Res 12, 1845-1855.

European Association for the Study of the Liver. Electronic address, e.e.e., 2016. EASL Clinical Practice Guidelines: Liver transplantation. J Hepatol 64, 433-485.

Fukudo, M., Yano, I., Fukatsu, S., Saito, H., Uemoto, S., Kiuchi, T., Tanaka, K., Inui, K., 2003. Forecasting of blood tacrolimus concentrations based on the Bayesian method in adult patients receiving living-donor liver transplantation. Clin Pharmacokinet 42, 1161-1178.

Goto, M., Masuda, S., Kiuchi, T., Ogura, Y., Oike, F., Okuda, M., Tanaka, K., Inui, K., 2004. CYP3A5*1-carrying graft liver reduces the concentration/oral dose ratio of tacrolimus in recipients of living-donor liver transplantation. Pharmacogenetics 14, 471-478.

Hao, J.C., Wang, W.T., Yan, L.N., Li, B., Wen, T.F., Yang, J.Y., Xu, M.Q., Zhao, J.C., Wei, Y.G., 2014. Effect of low-dose tacrolimus with mycophenolate mofetil on renal function following liver transplantation. World J Gastroenterol 20, 11356-11362.

Hesse, C.J., Baan, C.C., Balk, A.H., Metselaar, H.J., Weimar, W., van Gelder, T., 2002. Evaluation of the new EMIT enzyme immunoassay for the determination of whole-blood tacrolimus concentrations in kidney, heart, and liver transplant recipients. Transplant Proc 34, 2988-2990.

Horina, J.H., Wirnsberger, G.H., Kenner, L., Holzer, H., Krejs, G.J., 1993. Increased susceptibility for CsA-induced hepatotoxicity in kidney graft recipients with chronic viral hepatitis C. Transplantation 56, 1091-1094.

Ji, E., Choi, L., Suh, K.S., Cho, J.Y., Han, N., Oh, J.M., 2012. Combinational effect of intestinal and hepatic CYP3A5 genotypes on tacrolimus pharmacokinetics in recipients of living donor liver transplantation. Transplantation 94, 866-872.

Ji, E., Kim, M.G., Oh, J.M., 2018. CYP3A5 genotype-based model to predict tacrolimus dosage in the early postoperative period after living donor liver transplantation. Ther Clin Risk Manag 14, 2119-2126.

Laporte-Simitsidis, S., Girard, P., Mismetti, P., Chabaud, S., Decousus, H., Boissel, J.P., 2000. Inter-study variability in population pharmacokinetic meta-analysis: when and how to estimate it? J Pharm Sci 89, 155-167.

Lee, D.R., Ho, M.J., Jung, H.J., Cho, H.R., Park, J.S., Yoon, S.H., Choi, Y.S., Choi, Y.W., Oh, C.H., Kang, M.J., 2016. Enhanced dissolution and oral absorption of tacrolimus by supersaturable self-emulsifying drug delivery system. Int J Nanomedicine 11, 1109-1117.

Lee, J.Y., Hahn, H.J., Son, I.J., Suh, K.S., Yi, N.J., Oh, J.M., Shin, W.G., 2006. Factors affecting the apparent clearance of tacrolimus in Korean adult liver transplant recipients. Pharmacotherapy 26, 1069-1077.

LeGatt, D.F., Shalapay, C.E., Cheng, S.B., 2004. The EMIT 2000 tacrolimus assay: an application protocol for the Beckman Synchron LX20 PRO analyzer. Clin Biochem 37, 1022-1030.

Li, D., Lu, W., Zhu, J.Y., Gao, J., Lou, Y.Q., Zhang, G.L., 2007. Population pharmacokinetics of tacrolimus and CYP3A5, MDR1 and IL-10 polymorphisms in adult liver transplant patients. J Clin Pharm Ther 32, 505-515.

Lu, Y.X., Su, Q.H., Wu, K.H., Ren, Y.P., Li, L., Zhou, T.Y., Lu, W., 2015. A population pharmacokinetic study of tacrolimus in healthy Chinese volunteers and liver transplant patients. Acta Pharmacol Sin 36, 281-288.

Mao, J.J., Jiao, Z., Yun, H.Y., Zhao, C.Y., Chen, H.C., Qiu, X.Y., Zhong, M.K., 2018. External evaluation of population pharmacokinetic models for ciclosporin in adult renal transplant recipients. Br J Clin Pharmacol 84, 153-171.

McDougall, D.A.J., Martin, J., Playford, E.G., Green, B., 2016. The Impact of Model-Misspecification on Model Based Personalised Dosing. AAPS J 18, 1244-1253.

Mizuno, T., Emoto, C., Fukuda, T., Hammill, A.M., Adams, D.M., Vinks, A.A., 2017. Model-based precision dosing of sirolimus in pediatric patients with vascular anomalies. Eur J Pharm Sci 109S, S124-S131.

Mizuno, T., O'Brien, M.M., Vinks, A.A., 2019. Significant effect of infection and food intake on sirolimus pharmacokinetics and exposure in pediatric patients with acute lymphoblastic leukemia. Eur J Pharm Sci 128, 209-214.

Oteo, I., Lukas, J.C., Leal, N., Suarez, E., Valdivieso, A., Gastaca, M., Ortiz de Urbina, J., Calvo, R., 2013. Tacrolimus pharmacokinetics in the early post-liver transplantation period and clinical applicability via Bayesian prediction. Eur J Clin Pharmacol 69, 65-74.

Passey, C., Birnbaum, A.K., Brundage, R.C., Oetting, W.S., Israni, A.K., Jacobson, P.A., 2011. Dosing equation for tacrolimus using genetic variants and clinical factors. Br J Clin Pharmacol 72, 948-957.

Press, R.R., Ploeger, B.A., den Hartigh, J., van der Straaten, T., van Pelt, J., Danhof, M., de Fijter, J.W., Guchelaar, H.J., 2009. Explaining variability in tacrolimus pharmacokinetics to optimize early exposure in adult kidney transplant recipients. Ther Drug Monit 31, 187-197.

Sam, W.J., Aw, M., Quak, S.H., Lim, S.M., Charles, B.G., Chan, S.Y., Ho, P.C., 2000. Population pharmacokinetics of tacrolimus in Asian paediatric liver transplant patients. Br J Clin Pharmacol 50, 531-541.

Sam, W.J., Tham, L.S., Holmes, M.J., Aw, M., Quak, S.H., Lee, K.H., Lim, S.G., Prabhakaran, K., Chan, S.Y., Ho, P.C., 2006. Population pharmacokinetics of tacrolimus in whole blood and plasma in asian liver transplant patients. Clin Pharmacokinet 45, 59-75.

Sarhan, M.D., Osman, A.M.A., Mohamed, M.A., Abdelaziz, O., Serour, D.K., Mansour, D.A., Mogawer, S., Helmy, A.S., El-Shazli, M.A., Hosny, A.A., 2017. Biliary Complications in Recipients of Living-Donor Liver Transplant: A Single-Center Review of 120 Patients. Exp Clin Transplant 15, 648-657.

Sheiner, L.B., Beal, S.L., 1982. Bayesian individualization of pharmacokinetics: simple implementation and comparison with non-Bayesian methods. J Pharm Sci 71, 1344-1348.

Shimada, T., Terada, A., Yokogawa, K., Kaneko, H., Nomura, M., Kaji, K., Kaneko, S., Kobayashi, K., Miyamoto, K., 2002. Lowered blood concentration of tacrolimus and its recovery with changes in expression of CYP3A and P-glycoprotein after high-dose steroid therapy. Transplantation 74, 1419-1424.

Shimomura, M., Masuda, S., Saito, H., Sakamoto, S., Uemoto, S., Tanaka, K., Inui, K., 2002. Roles of the jejunum and ileum in the first-pass effect as absorptive barriers for orally administered tacrolimus. J Surg Res 103, 215-222.

Shuker, N., van Gelder, T., Hesselink, D.A., 2015. Intra-patient variability in tacrolimus exposure: causes, consequences for clinical management. Transplant Rev (Orlando) 29, 78-84.

Siekierka, J.J., Staruch, M.J., Hung, S.H., Sigal, N.H., 1989. FK-506, a potent novel immunosuppressive agent, binds to a cytosolic protein which is distinct from the cyclosporin A-binding protein, cyclophilin. J Immunol 143, 1580-1583.

Staatz, C.E., Tett, S.E., 2004. Clinical pharmacokinetics and pharmacodynamics of tacrolimus in solid organ transplantation. Clin Pharmacokinet 43, 623-653.

Staatz, C.E., Willis, C., Taylor, P.J., Lynch, S.V., Tett, S.E., 2003. Toward better outcomes with tacrolimus therapy: population pharmacokinetics and individualized dosage prediction in adult liver transplantation. Liver Transpl 9, 130-137.

Tamura, S., Ohike, A., Ibuki, R., Amidon, G.L., Yamashita, S., 2002. Tacrolimus is a class II low-solubility high-permeability drug: the effect of P-glycoprotein efflux on regional permeability of tacrolimus in rats. J Pharm Sci 91, 719-729.

Taner, C.B., Bulatao, I.G., Perry, D.K., Sibulesky, L., Willingham, D.L., Kramer, D.J., Nguyen, J.H., 2012. Asystole to cross-clamp period predicts development of biliary complications in liver transplantation using donation after cardiac death donors. Transpl Int 25, 838-846.

Tauzin, M., Treluyer, J.M., Nabbout, R., Billette de Villemeur, T., Desguerre, I., Aboura, R., Gana, I., Zheng, Y., Benaboud, S., Bouazza, N., Chenevier-Gobeaux, C., Freihuber, C., Hirt, D., 2019. Simulations of Valproate Doses Based on an External Evaluation of Pediatric Population Pharmacokinetic Models. J Clin Pharmacol 59, 406-417.

Tubic, M., Wagner, D., Spahn-Langguth, H., Bolger, M.B., Langguth, P., 2006. In silico modeling of non-linear drug absorption for the P-gp substrate talinolol and of consequences for the resulting pharmacodynamic effect. Pharm Res 23, 1712-1720.

Vadcharavivad, S., Praisuwan, S., Techawathanawanna, N., Treyaprasert, W., Avihingsanon, Y., 2016. Population pharmacokinetics of tacrolimus in Thai kidney transplant patients: comparison with similar data from other populations. J Clin Pharm Ther 41, 310-328.

Vanhove, T., Annaert, P., Kuypers, D.R., 2016a. Clinical determinants of calcineurin inhibitor disposition: a mechanistic review. Drug Metab Rev 48, 88-112.

Vanhove, T., de Jonge, H., de Loor, H., Annaert, P., Diczfalusy, U., Kuypers, D.R., 2016b. Comparative performance of oral midazolam clearance and plasma 4beta-hydroxycholesterol to explain interindividual variability in tacrolimus clearance. Br J Clin Pharmacol 82, 1539-1549.

Wallemacq, P., Armstrong, V.W., Brunet, M., Haufroid, V., Holt, D.W., Johnston, A., Kuypers, D., Le Meur, Y., Marquet, P., Oellerich, M., Thervet, E., Toenshoff, B., Undre, N., Weber, L.T., Westley, I.S., Mourad, M., 2009. Opportunities to optimize tacrolimus therapy in solid organ transplantation: report of the European consensus conference. Ther Drug Monit 31, 139-152.

Wallin, J.E., Bergstrand, M., Wilczek, H.E., Nydert, P.S., Karlsson, M.O., Staatz, C.E., 2011. Population pharmacokinetics of tacrolimus in pediatric liver transplantation: early posttransplantation clearance. Ther Drug Monit 33, 663-672.

Yu, M., Liu, M., Zhang, W., Ming, Y., 2018. Pharmacokinetics, Pharmacodynamics and Pharmacogenetics of Tacrolimus in Kidney Transplantation. Curr Drug Metab 19, 513-522.

Zahir, H., McLachlan, A.J., Nelson, A., McCaughan, G., Gleeson, M., Akhlaghi, F., 2005. Population pharmacokinetic estimation of tacrolimus apparent clearance in adult liver transplant recipients. Ther Drug Monit 27, 422-430.

Zhang, H.X., Sheng, C.C., Liu, L.S., Luo, B., Fu, Q., Zhao, Q., Li, J., Liu, Y.F., Deng, R.H., Jiao, Z., Wang, C.X., 2019. Systematic external evaluation of published population pharmacokinetic models of mycophenolate mofetil in adult kidney transplant recipients co-administered with tacrolimus. Br J Clin Pharmacol 85, 746-761.

Zhang, X.Q., Wang, Z.W., Fan, J.W., Li, Y.P., Jiao, Z., Gao, J.W., Peng, Z.H., Liu, G.L., 2012. The impact of sulfonylureas on tacrolimus apparent clearance revealed by a population pharmacokinetics analysis in Chinese adult liver-transplant patients. Ther Drug Monit 34, 126-133.

Zhao, C.Y., Jiao, Z., Mao, J.J., Qiu, X.Y., 2016. External evaluation of published population pharmacokinetic models of tacrolimus in adult renal transplant recipients. Br J Clin Pharmacol 81, 891-907.

Zhu, L., Wang, H., Sun, X., Rao, W., Qu, W., Zhang, Y., Sun, L., 2014. The Population Pharmacokinetic Models of Tacrolimus in Chinese Adult Liver Transplantation Patients. J Pharm (Cairo) 2014, 713650.

Zhu, L., Yang, J., Zhang, Y., Jing, Y., Zhang, Y., Li, G., 2015. Effects of CYP3A5 genotypes, ABCB1 C3435T and G2677T/A polymorphism on pharmacokinetics of Tacrolimus in Chinese adult liver transplant patients. Xenobiotica 45, 840-846.

# Tables

**Table 1** Summary of published population pharmacokinetic studies of tacrolimus in adult liver transplant recipients

| Study (publication year) | Country (Single/ multiple sites) | Number of patients (Male/Female) | Sampling schedule (Number of samples) | Postoperative time mean±SD / median(range) | Bioassay | Structural model | PK parameters and formulas | | BSV% (BOV%) | RUV |
|---|---|---|---|---|---|---|---|---|---|---|
| [1] Fukudo et al.(2003) | Japan (Single) | 35(18/17) | $C_0$ (824) | (0-28) days | MEIA | 1CMT | CL | $0.743 + 0.0157 \times POD$ $\times (0.792, \text{if TBIL} > 42.75)$ $\times (0.810, \text{if SCR} > 88.4)$ $\times GHW/600$ | 60.0 [i] | 2.57 ng ml$^{-1}$ |
| | | | | | | | $V_d$ | $1.64 \times WT$ | 35.4 | |
| | | | | | | | F | 0.0732 | 71.2 [i] | |
| [2] Staatz et al.(2003) | Australia (multiple) | 68(48/20) | SS[a] (1742) | 323 ± 370 (6-2115) days | LC-MS/MS | 1CMT | CL/F | 29.6 (if AST < 70U/L) or 24 (if AST > 70U/L) | 43 | 3.3 ng ml$^{-1}$ |
| | | | | | | | $V_d$/F | $601 \times (WT/72.1)$ | 93 | |
| | | | | | | | Ka | 4.48 (fixed) | / | |
| [3] Dansirikul et al.(2004) | Australia (multiple) | 31(23/8) | SS[b] (not available) totally 54 PK profiles | 420 ± 530 (11-1886) days | LC-MS/MS | 2CMT | CL/F | 27.0 | 43.4 (16.4) [i] | 14.4% |
| | | | | | | | $V_C$/F | 388.0 | 38.7 (61.2) [i] | |
| | | | | | | | Q/F | 54.5 | / | |
| | | | | | | | $V_P$/F | 2200 | / | |
| | | | | | | | Ka | 4 | 69.6 | |
| [4] Antignac et al.(2005) | France (single) | 37(26/11) | $C_0$ (824) | 29 ± 12 (11-66) days | MEIA | 1CMT | $CL_{max}$ | $36 \times (ALB/38)^{-0.64}$ | 43.6 [i] | 3.07 ng ml$^{-1}$ |
| | | | | | | | $TCL_{50}$ | $6.3 \times (AST/46)^{0.28}$ | 33.2 | |
| | | | | | | | CL | $CL_{max} \times POD^{4.9}/(TCL_{50}^{4.9} + POD^{4.9})$ | / | |
| | | | | | | | $V_d$ | 1870 | 49.0 [i] | |
| | | | | | | | Ka | 4.48 (fixed) | / | |
| [5] Zahir et al.(2005) | Australia (single) | 67(45/22) | $C_0$ (694) | 40.2 ± 17.3 (14-94) days | MEIA | 1CMT | CL/F | 21.3 + (9.8, if HCT < 35) + (3.4, if ALB < 35) - (2.1, if concomitant diltiazem) - (7.4, if concomitant fluconazole) | 31.6 | 24.3% |
| | | | | | | | $V_d$/F | 314 | / | |
| | | | | | | | Ka | 4.5 (fixed) | / | |

*(Continues)*

**Table 1** continued

| Study (publication year) | Country (Single/ multiple sites) | Number of patients (Male/Female) | Sampling schedule (Number of samples) | Postoperative time mean±SD / median(range) | Bioassay | Structural model | PK parameters and formulas | | BSV% (BOV%) | RUV |
|---|---|---|---|---|---|---|---|---|---|---|
| [6] Lee et al.(2006) | South Korea (single) | 35(25/10) | $C_0$ (1251) | 133.03 ± 74.08 (24-308) days | MEIA | 1CMT | CL/F | 0.36 + 2.01/POD (if POD ≤ 35) × TBIL $^{-0.23}$ (if TBIL > 20.52) × (0.49, if POD ≤ 3) × (0.75, if INR > 1.4) × (0.86, if GRWR ≤ 1.25%) × WT | 35.35 | 3.14 ng ml$^{-1}$ |
| | | | | | | | $V_d/F$ | 568 | 68.12 | |
| [7] Sam et al.(2006) | Singapore (single) | 31 (23/8) | IS$^c$ (213) | Not available | HPLC-MS/MS | 1CMT | CL/F | 14.1 + 0.237 × (WT-55) - (2.93, if ALP ≥ 200) - 0.801 × (SCR - 60) | 65.7 | 34.8% |
| | | | | | | | $V_d/F$ | 217 - 7.83 × (HCT-31.1) + 179 × (HT-1.61) | 63.8 | |
| | | | | | | | Ka | 2.08 | / | |
| [8] Li et al.(2007) | China (single) | 72(60/12) | $C_0$ (703) | 13.5 (1-88.5) days | MEIA | 1CMT | CL/F | 15.9 - 1.88 × [(1, if 25.7 < TBIL ≤ 51.4) or (2, if 51.4 < TBIL ≤ 77.1) or (3, 77.1 < TBIL ≤ 128.5) or (4, TBIL > 128.5) + (7.65, if donor CYP3A5*1) + (7.00, if recipient CYP3A5*1) | 31.2 | 2.81 ng ml$^{-1}$ |
| | | | | | | | $V_d/F$ | 620 | 55.0 | |
| | | | | | | | Ka | 4.48 (fixed) | / | |
| [9] Blanchet et al.(2008) | France (single) | 14(11/3) | IS$^d$(198) | (6-97) days | EMIT | 2CMT | CL/F | 2.85 × (0.36, if a whole/split graft) × (1.026, if coagulation factor V) | 23 | 11% |
| | | | | | | | $V_C/F$ | 87 | 44 | |
| | | | | | | | Q/F | 22 | 39 | |
| | | | | | | | $V_P/F$ | 1290 | 36 | |
| | | | | | | | Ka | 4.03 | 44 | |

**Table 1** continued

| Study (publication year) | Country (Single/ multiple sites) | Number of patients (Male/Female) | Sampling schedule (Number of samples) | Postoperative time mean±SD / median(range) | Bioassay | Structural model | PK parameters and formulas | | BSV% (BOV%) | RUV |
|---|---|---|---|---|---|---|---|---|---|---|
| [10] Zhang et al.(2012) | China (single) | 262 (226/36) | $C_0$ (3703) | 85.4±211.5 37.5 (2-1941) days | MEIA | 1CMT | CL/F | $20.9 \times (DD/4)^{0.582}$ $\times (HCT/35.4)^{0.418}$ $\times (TP/69.1)^{0.780}$ $\times (0.841,$ if concomitant sulfonylureas) | 23.8 | 33.6% 0.96 ng ml$^{-1}$ |
| | | | | | | | $V_d/F$ | $808 \times (HCT/35.4)^{1.52} \times (TP/69.1)^{1.81}$ | 70.4 | |
| | | | | | | | Ka | 4.0 (fixed) | / | |
| [11] Oteo et al.(2013) | Spain (single) | 75(not available) | $C_0$ (335) | (0-15) days | MEIA | 1CMT | CL/F | 11.10 (if POD ≤ 3 & AST< 500) or 8.04 (if POD ≤ 3 & AST ≥ 500) or 17.80 (if 4 ≤ POD ≤ 15 & ALB ≥ 25 & HCT ≥ 28) or 24.50 (if 4 ≤ POD ≤ 15 & ALB < 25 & HCT < 28) | 45.93$^e$ 36.74$^f$ | 28.04% |
| | | | | | | | $V_d/F$ | 328 (if POD ≤ 3) or 568 (if 4 ≤ POD ≤ 15) | 52.15$^e$ 20.20$^f$ | |
| | | | | | | | Ka | 4.48 (fixed) | / | |
| [12] Zhu et al.(2014) | China (single) | 47 (27/20) | IS$^g$ (435) | 20.71±18.04 14 (2-85) days | MEIA | 2CMT | CL/F | $11.2 \times DD \times POD^{0.127}$ | 16.2 | 26.54% |
| | | | | | | | $V_C/F$ | 406 | 163 | |
| | | | | | | | Q/F | 57.3 | 19.7 | |
| | | | | | | | $V_P/F$ | 503 | 199 | |
| | | | | | | | Ka | 0.723 | 74.3 | |
| [13] Lu et al.(2015) | China (single) | 112 (86/26) | $C_0$ (1100) | 19.38 ± 17.75 (2-137) days | MEIA | 2CMT | CL/F | $32.8 \times 0.562 \times [EXP(ALT/40) \times (-0.0237)]$ | 46.6 | 39.8% |
| | | | | | | | $V_C/F$ | 22.7 | 57.3 | |
| | | | | | | | Q/F | 76.3 | 46 | |
| | | | | | | | $V_P/F$ | 916 (fixed) | 93.5 | |
| | | | | | | | Ka | 0.419 | / | |
| | | | | | | | Tlag | 0.404 | / | |

**Table 1** continued

| Study (publication year) | Country (Single/multiple sites) | Number of patients (Male/Female) | Sampling schedule (Number of samples) | Postoperative time mean±SD / median(range) | Bioassay | Structural model | PK parameters and formulas | | BSV% (BOV%) | RUV |
|---|---|---|---|---|---|---|---|---|---|---|
| [14] Zhu et al.(2015) | China (single) | 95(73/22) | $C_0$ (2285) | 39.5 (1-341) days | MEIA | 1CMT | CL/F | $17.6 \times (POD/40.36)^{0.205} \times (BUN/11.86)^{-0.116}$ $\times (ALP/149.77)^{0.165}$ $\times (TBIL/100.22)^{-0.142}$ $\times (HCT/99.09)^{-0.789}$ $\times (1.661,$ if CYP3A5*1 recipient$)$ | 53.9 | 28.40% 0.606 ng ml$^{-1}$ |
| | | | | | | | $V_d/F$ | $225 \times POD^{0.852} \times (HB/99.09)^{-0.813}$ | 68 | |
| | | | | | | | Ka | 4.48 (fixed) | / | |
| [15] Chen et al.(2017) | China (single) | 153 (125/28) | $C_0$ (1234) + IS$^h$ (470) | 22.8 ± 6.78 (3-89) days | MEIA LC-MS/MS | 2CMT | CL/F | $21.9 \times EXP(0.0102 \times POD)$ $\times EXP(0.258 \times CCR/113)$ $\times EXP(-0.148)$ (if ABCB1 3435CT recipient) or $EXP(-0.296)$ ( if ABCB1 3435TT recipient) | 36.3 | 33.3% |
| | | | | | | | $V_C/F$ | $284 \times EXP(-0.125)$ ( if ABCB1 3435CT recipient) or $EXP(-0.25)$ (if ABCB1 3435TT recipient) | 89.4 | |
| | | | | | | | Q/F | 62.1 | / | |
| | | | | | | | $V_P/F$ | 710 | / | |
| | | | | | | | Ka | 0.55 | 56.7 | |
| | | | | | | | Tlag | 1.96 | / | |
| [16] Ji et al.(2018) | Korea (single) | 58 (46/12) | $C_0$ (605) | (0-14) days | MEIA | 1CMT | CL/F | $6.33 \times POD^{0.257}$ $\times 2.314$ (if CYP3A5*1 recipient grafted from CYP3A5*1 donor) $\times 1.523$ (if CYP3A5*1 recipient grafted from CYP3A5 *3/*3 donor) | 34.2 | 42.70% 0.915 ng ml$^{-1}$ |
| | | | | | | | $V_d/F$ | $465 \times POD^{0.322}$ | 45.5 | |
| | | | | | | | Ka | 4.48 (fixed) | / | |

ALB, albumin (g l$^{-1}$); ALP, alkaline phosphatase (U l$^{-1}$); ALT, alanine aminotransferase (U l$^{-1}$); AST, aspartate transferase (U l$^{-1}$); BOV, between occasion variability; BSV, between subject variability; BUN, blood uric nitrogen (mmol l$^{-1}$) ; WT, body weight (kg); $C_0$, blood trough concentration; CL, clearance (l h$^{-1}$); CL/F, apparent clearance (l h$^{-1}$); CCR, creatinine clearance calculated by the Cockcroft-Gault formula (ml min$^{-1}$); CMT, compartment; CYP3A5*1, cytochrome P450 3A5 expresser (*1/*1 or *1/*3); DD, tacrolimus daily dose (mg day$^{-1}$); DDWT, tacrolimus daily dose per kilogram of body weight (mg day$^{-1}$ kg$^{-1}$); EMIT, enzyme multiplied immunoassay technique; F, bioavailability; GHW, grafted hepatic weight (g); GRWR, graft:recipient weight ratio (%); HB, hemoglobin (g l$^{-1}$); HCT, haematocrit (%); HPLC, high performance liquid chromatography; HT, body height (m); INR, international normalized ratio; IS, intensive sampling; Ka, absorption rate constant (h$^{-1}$); LC–MS/MS, liquid chromatography tandem-mass spectrometry; MEIA, microparticle enzyme immunoassay; POD, postoperative days (day); Q, inter-compartmental clearance (l h$^{-1}$); Q/F, apparent inter-compartmental clearance (l h$^{-1}$); RUV, residual unexplained variability; SCR, serum creatinine (µmol l⁻¹); SD, standard deviation; SS, sparse sampling; TBIL, Total bilirubin (µmol l⁻¹); Tlag, absorption lag time (h); TP, total protein (g l⁻¹); $V_C$, volume of distribution of central compartment (l); $V_C/F$, apparent volume of distribution of central compartment (l); $V_P$, volume of distribution of peripheral compartment (l); $V_P/F$, apparent volume of distribution of peripheral compartment (l);

[a] Intensive samples were collected at 0, 1, 2, 4, and 6 or 8 h postdose and repeated on a second occasion (generally 1 to 4 weeks after the first collection day).

[b] Intensive samples were collected at 0, 1, 2, 4, and 6 h postdose.

[c] Intensive samples were collected predose and at specific time-points after an oral dose of tacrolimus. But, details and the time points of sampling were not available.

[d] Intensive samples were collected predose and at 2, 3, 4, 6, and 9 h postdose on day 8 (±2), day 21 (±3), and day 90 (±7) post-transplantation.

[e] Between subject variability for CL/F in the 0-3 day post-transplantation period.

[f] Between subject variability for Vd/F in the 4-15 day post-transplantation period.

[g] Most of the intensive samples were collected predose and at 0.3, 1, 1.5, 2, 4, 6, 8, and 12 h postdose while some were collected only before administration in steady-state conditions.

[h] Intensive samples (28 patients) were collected at 0, 1, 1.5, 2, 2.5, 3, 4, 6, 8 and 12 h postdose at week 1 and 3 post-transplantation.

[i] correlation coefficient: CL ~ F: 0.770; CL/F ~ $V_c/F$: 0.583; $CL_{max}$ ~ $V_d$: 0.55;

**Table 2** Characteristics of external evaluation dataset

| Characteristics | Number or mean ± SD | Median (range) |
|---|---|---|
| No. of patients (Male/Female) | 84 (69/15) | / |
| No. of samples | 572 | / |
| Primary disease | | |
|     virus related cirrhosis | | |
|         hepatitis B virus [a] | 61 | / |
|         hepatitis C virus [a] | 3 | / |
|         hepatitis E virus [a] | 1 | / |
|     hepatocellular carcinoma [a] | 13 | / |
|     alcohol cirrhosis [a] | 4 | / |
|     autoimmune hepatitis [a] | 2 | / |
| Age (years) | 51±9.97 | 51(17-74) |
| Body weight, WT (kg) | 63.24±11.14 | 62(40-100) |
| Body height (m) | 1.69±0.07 | 1.70(1.54-1.80) |
| Grafted hepatic weight, GHW (g) | 1299.25±218.51 | 1300(603-2100) |
| Graft: recipient weight ratio, GRWR (%) | 2.14±0.51 | 2.03(1.04-3.58) |
| Tacrolimus daily dose (mg day$^{-1}$) | 3.13±1.61 | 3.00(0.25-8.00) |
| Tacrolimus trough concentration (ng ml$^{-1}$) | 6.78±2.94 | 6.59(1.00-23.11) |
| Postoperative days, POD (day) | 13.8±8.5 | 12(4-50) |
| Hemoglobin, HB (g l$^{-1}$) | 104.64±16.7 | 104(61-159) |
| Haematocrit, HCT (%) | 31.28±4.99 | 31(16.6-53.6) |
| Albumin, ALB (g l$^{-1}$) | 37.28±4.05 | 37(20-51) |
| Total protein, TP (g l$^{-1}$) | 60.67±7.03 | 60(40-91) |
| Alanine aminotransferase, ALT (U l$^{-1}$) | 105.86±141.41 | 49.5(4-1043) |
| Aspartate aminotransferase, AST (U l$^{-1}$) | 64.03±78.29 | 42(10-886) |
| Alkaline phosphatase, ALP (U l$^{-1}$) | 238.71±231.13 | 167.5(44-2021) |
| Total bilirubin, TBIL (μmol l$^{-1}$) | 89.89±78.57 | 66.25(5.1-458.8) |
| International normalized ratio, INR | 1.24±0.19 | 1.24(0.91-2.77) |
| Blood uric nitrogen, BUN (mmol l$^{-1}$) | 8.95±4.9 | 7.85(1.5-36) |
| Serum creatinine, SCR (μmol l$^{-1}$) | 61.96±29.82 | 57(25-400) |
| Creatinine clearance, CCR (ml min$^{-1}$) [b] | 123.25±44.88 | 122.22(23.36-319.46) |
| Concomitant medication | | |
|     Diltiazem [a] | 1 | / |
|     Fluconazole [a] | 7 | / |
|     Sulfonylureas [a] | 0 | / |
|     mycophenolate mofetil [a] | 5 | / |

SD, standard deviation

[a] Data are expressed as number of recipients (samples)

[b] Calculated from serum creatinine using the Cockcroft–Gault formula: CCR= [140 – age (years)] × weight (kg) / [0.818 × SCR (μmol l$^{-1}$)] × (0.85, if female)

Table 3 Allele frequencies of genetic polymorphisms in CYP3A5 genes of external dataset

| Single nucleotide polymorphisms | Number of recipients | Frequency (%) |
|---|---|---|
| **Recipients** | | |
| CYP3A5*3 (A6986G, rs776746) | | |
|     AA (*1/*1) | 5 | 5.95 |
|     GA (*1/*3) | 35 | 41.67 |
|     GG (*3/*3) | 44 | 52.38 |
| **Donors** | | |
| CYP3A5*3 (A6986G, rs776746) | | |
|     AA (*1/*1) | 4 | 4.76 |
|     GA (*1/*3) | 39 | 46.43 |
|     GG (*3/*3) | 41 | 48.81 |

The allele frequencies are found to be in Hardy–Weinberg equilibrium ($P > 0.05$)

**Figure legends**

**Figure 1** Box plots of prediction error (PE%) for 16 published population pharmacokinetic models. Black solid, grey, and dark-green dashed lines are reference lines indicating PE of 0%, ± 20%, and ± 30%, respectively.

**Figure 2** Box plots of individual prediction error (IPE%) with Bayesian forecasting for 16 published population pharmacokinetic models in different scenarios (0 represents predictions with no prior observation; 1–4 represent predictions with one to four prior observations, respectively). In scenario *n*, prior *n* observations were used to estimate the individual prediction, and it was then compared with the corresponding observation.

**Figure 3** Box plots of prediction error (PE%) for investigated structural models. Black solid, grey, and dark-green dashed lines are reference lines indicating PE of 0%, ± 20%, and ± 30%, respectively. 1CMT, one-compartment model; 2CMT two-compartment model; MM, Michaelis–Menten model.

**Figure 4** Box plots of individual prediction error (IPE%) with Bayesian forecasting for investigated structural models in different scenarios (0 represents predictions with no prior observation; 1–4 represent predictions with one to four prior observations, respectively). 1CMT, one-compartment model; 2CMT two-compartment model; MM, Michaelis–Menten model.

**Figure 1**

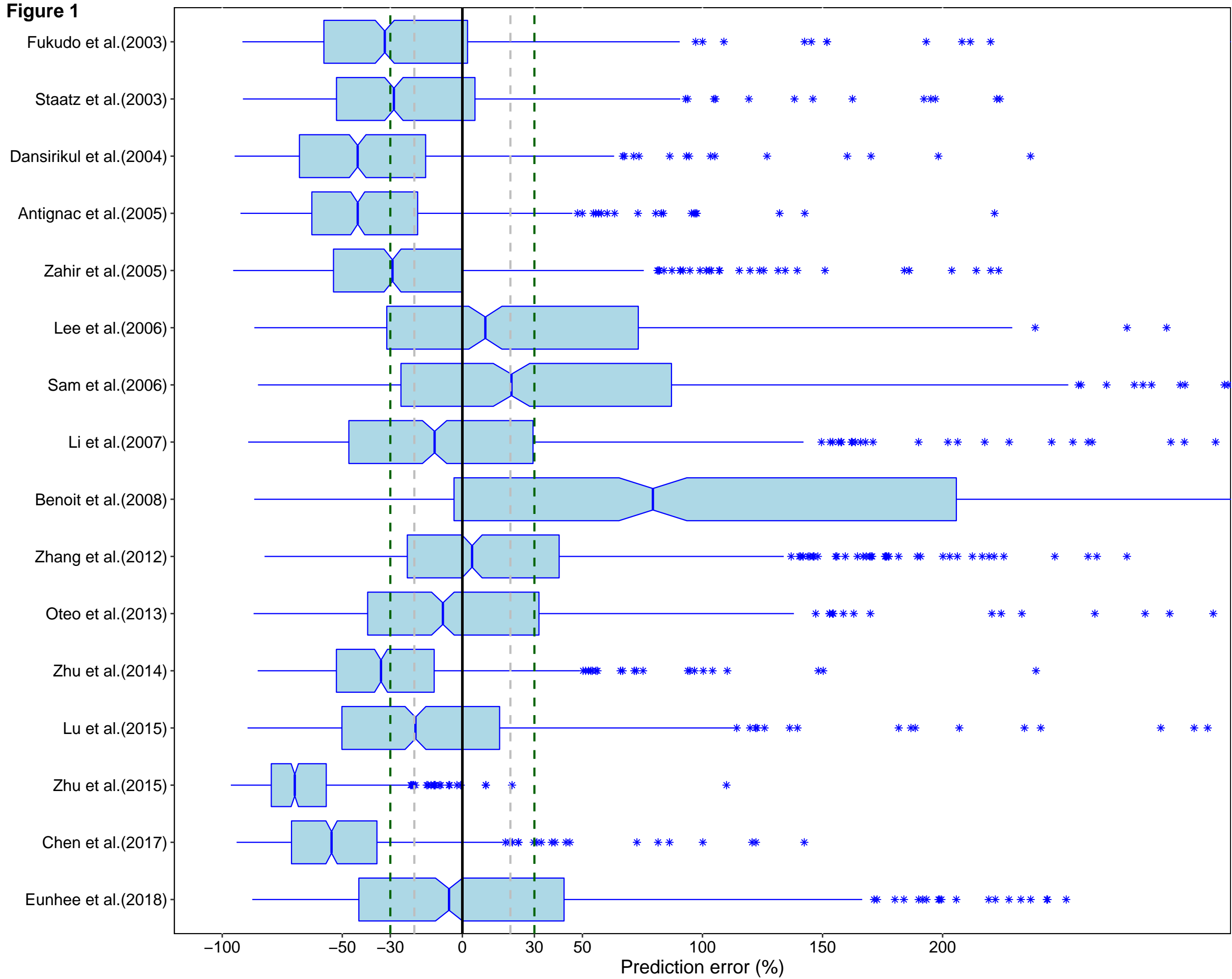

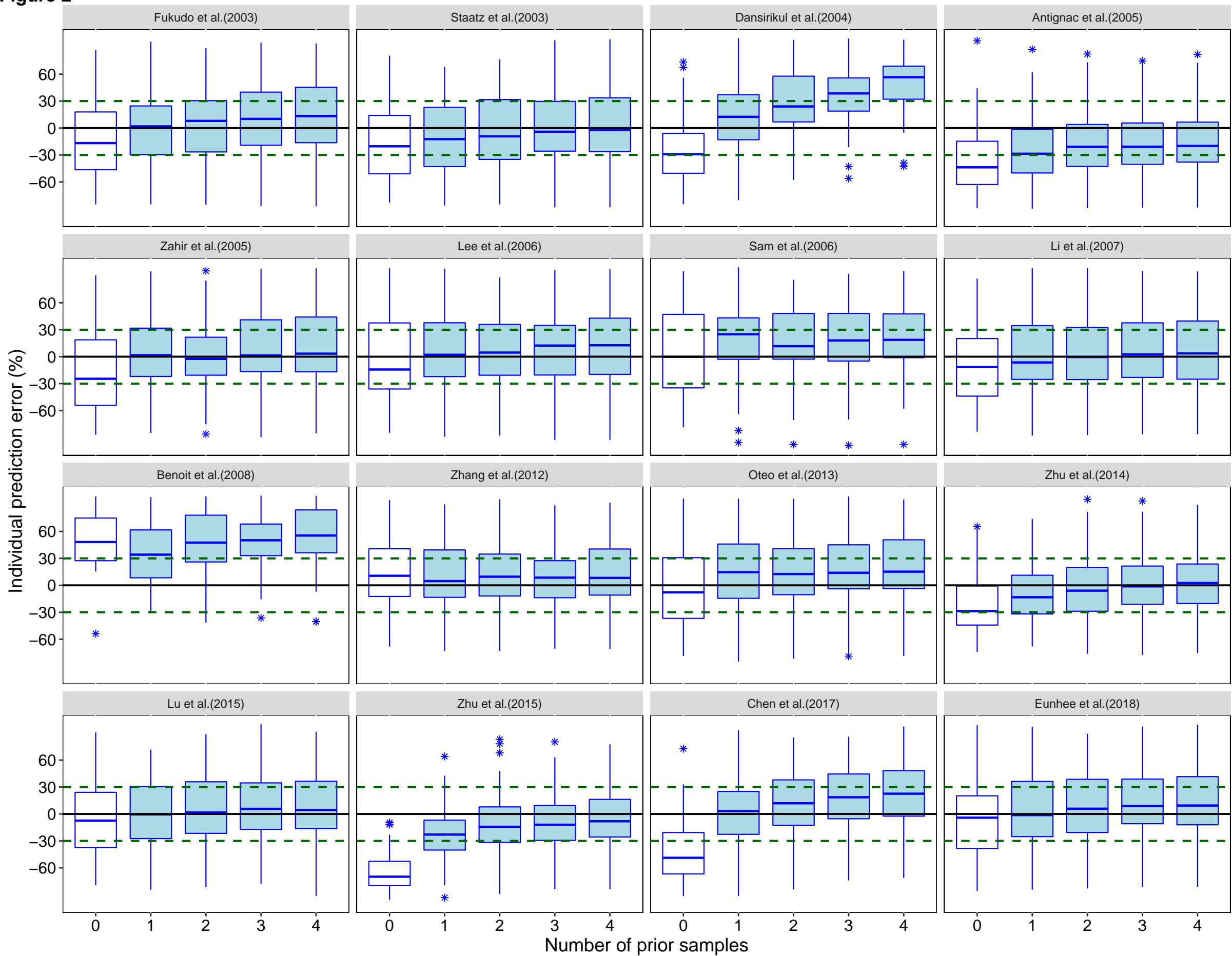

**Figure 3**

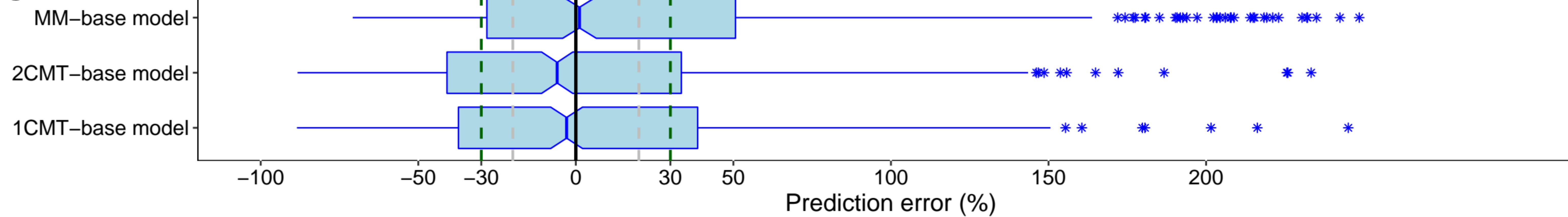

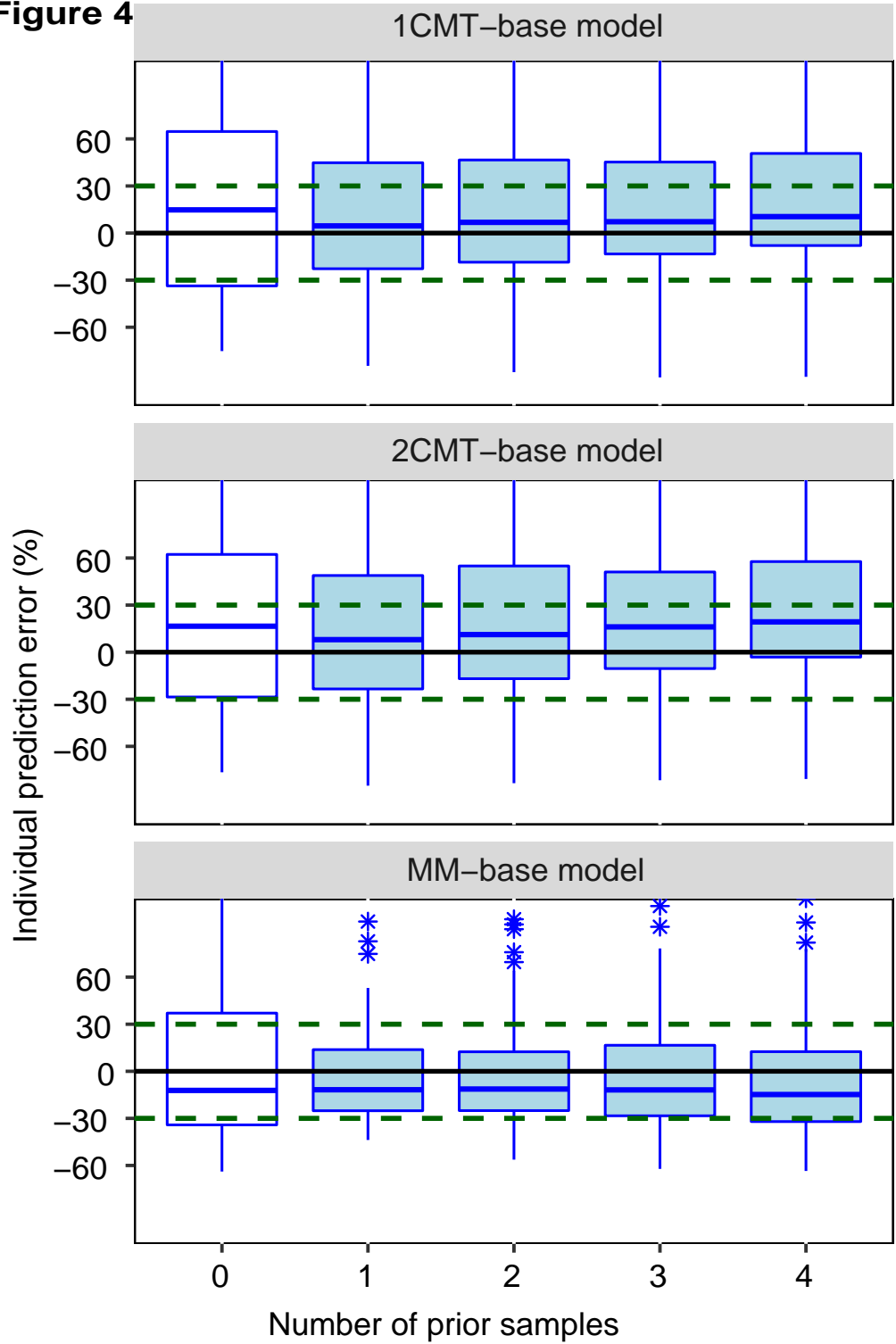

**Figure 4**

*Supporting information*

Additional Supporting Information may be found in the online version of this article at the publisher's web-site.

*Appendix S1*

Genotyping of *CYP3A5*3* single-nucleotide polymorphisms

*Appendix S2*

Detailed process of literature search and selection

*Appendix S3*

Summarized information of all the identified studies

*Table S1*

Results of prediction-based diagnostics

*Table S2*

Results of normalized prediction distribution error (NPDE) diagnostics

*Table S3*

Estimated parameters of investigated structural models

*Table S4*

Results of Bayesian forecasting

*Figure S1*

Population characteristics. (A) Histograms from different countries. Percentages of studies by sites where ethnicity of the patients has been reported in the published articles are indicated in the graph. (B) Histograms of studies based on the number of recipients included in the analysis. Number of recipients was categorized in 3 groups (below 50, above 100, and in between). (C) Histograms of studies based on the bioassay method performed to measure tacrolimus concentration. Bioassay was categorized in three methods including LC-MS/MS, EMIT, and MEIA. Percentages of each specific method are indicated for each category. (D) Frequencies of studies based on time posttransplant of follow-up. Postoperative time was categorized in different periods: first 2 weeks, first 1, 3, 6, and 12 months after transplant, after months posttransplant. EMIT, enzyme multiplied immunoassay technique; LC, liquid chromatography; MEIA, microparticle enzyme immunoassay; MS, mass spectrum; NR, not reported.

*Figure S2*

Prediction- and variability-corrected visual predictive check (pvcVPC) plot of the candidated models. **(A)** plotted with logarithmic scale for y-axis and linear scale for x-axis, **(B)** plotted with linear scale for both x- and y-axis. The red solid line represents the prediction- and variability-corrected median observed concentration, and the semitransparent red field represents the simulation-based 95% confidence intervals (CIs) for the median. The corrected observed 5$^{th}$ and 95$^{th}$ percentiles are shown with red dashed lines, and the simulation-based 95% CIs for the corresponding model predicted percentiles are presented with the semitransparent blue fields. The prediction- and variability-corrected observations are represented by blue dots.

*Figure S3*

Normalized prediction distribution error (NPDE) plots of the 16 candiated models. **(A)** Quantile-quantile plot of the distribution of the NPDE against the theoretical distribution (semitransparent blue fields), **(B)** Histogram of the distribution of the NPDE against the theoretical distribution (semitransparent blue fields), **(C)** NPDE vs. postoperative time (days), **(D)** NPDE vs. predicted concentrations. In plot C and D, the red solid lines represent the median NPDE of the observations, and semitransparent red fields represent the simulation-based 95% confidence intervals (CIs) for the median. Blue solid lines represent the NPDE of the observed 5$^{th}$ and 95$^{th}$ percentiles, and semitransparent blue fields represent the simulation-based 95% CIs for the corresponding model-predicted percentiles. The NPDE of the observations are represented by blue dots.

*Appendix S1. Genotyping of CYP3A5*3 single-nucleotide polymorphisms.*

DNA was extracted from the whole blood of both liver transplant receipients and their corresponding donors using the TIANamp Blood DNA Kit (Tiangen Biotech Co. Ltd, Beijing, China). The polymerase chain reaction (PCR) was applied to amplify the variant alleles using ABI Veriti 96-Well PCR (Applied Biosystems, Foster City, CA, USA). The volume of amplification reaction was 20 μl, containing 10 μl 2×Taq Master Mix, 7 μl of distilled water, 1 μl of each primer (10 pmol μl$^{-1}$), and 1 μl DNA template (20 ng μl$^{-1}$). The sequences of forward (F) and reverse (R) primers were listed below.

| The sequences of forward (F) and reverse (R) primers for the CYP3A5 genotyping SNPs | | |
|---|---|---|
| Gene | Primer | $T_m$ |
| CYP3A5*3 (rs776746) | F: 5' CATTTAGTCCTTGTGAGCACTTGAT 3' | 59.9 ℃ |
| | R: 5' TAGCACTGTTCTGATCACGTCG 3' | 58.8 ℃ |

The PCR conditions were as follows: a denaturation at 95 °C for 3 minutes, then 35 cycles of denaturation at 94 °C for 30 seconds, annealing at 55 °C for 25 seconds, and elongation at 72 °C for 30 seconds, followed by a final extension at 72 °C for 5 minutes. The amplified DNA was purified and genotypes were determined by direct sequencing using ABI PRISM 3730XL Sequence Detection System (Applied Biosystems, Foster City, CA, USA). Tm, melting temperature.

*Appendix S2: Detailed process of literature search and selection*

**Search terms:**

((liver transplant*) OR (hepatic transplant*) OR (liver graft*) OR (hepatic graft*)) AND ((population pharmacokinetic*) OR (non-linear mixed effect) OR NONMEM) AND (tacrolimus OR FK506 OR Prograf OR Advagraf), with an asterisk (*) used as a wildcard.

**Inclusion criteria:**

(1) Original population pharmacokinetic studies on adult liver transplant recipients treated with oral tacrolimus of the immediate-release formulation.
(2) Population pharmacokinetic modeling approach restricted to nonlinear mixed-effects analyses.
(3) Language: English.

**Exclusion criteria:**

- Receipients were not treated with oral tacrolimus of immediate-release formulation .
- Datasets were overlapped or studies were duplicated.
- Required covariates were unavailable in the evaluation dataset.
- Model details were not available for external evaluation.

*Flow diagram of literature selection process.*

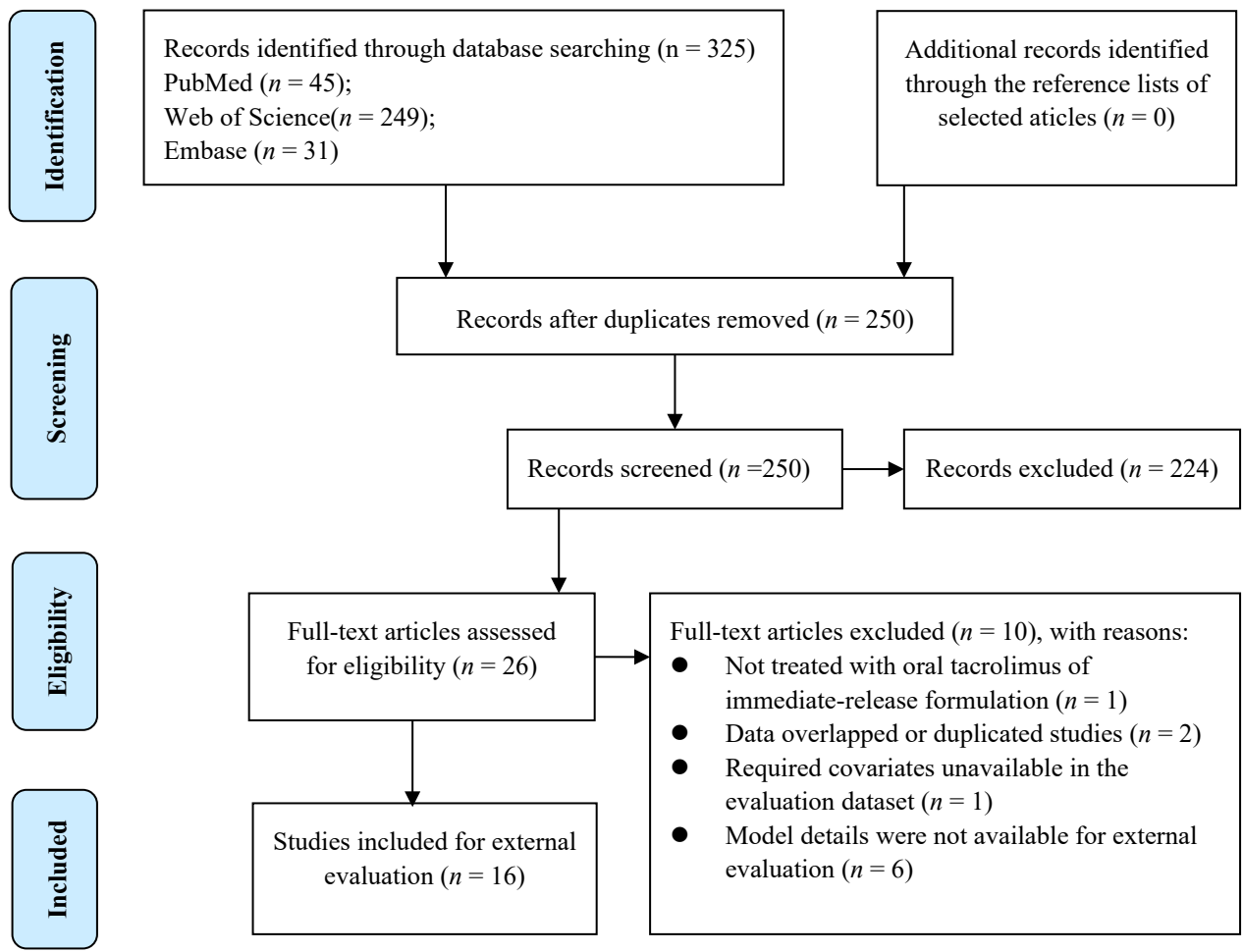

**Lists of the excluded studies (*n* = 10):**

(1) Not treated with oral tacrolimus of immediate-release formulation (*n* = 1)

- Receipients in the study by Moes, D. J. et al.[1] were treated with immunosuppressive therapy based on once-daily tacrolimus.

(2) Data or cohort overlapping (*n* = 2)

- Studies by Fukatsu et al. [2] and Fukudo et al. [3] had data overlap, thus the former one was excluded since the latter one was more recent.
- Studies by Valdivieso et al. [4] and Oteo et al. [5] had data overlapped and then the former one was excluded.

(3) Required covariates unavailable in the evaluation dataset (*n* = 1)

- The evaluation dataset missed the mRNA level of intestinal multidrug resistance-associated protein 1, which was the required covariates in the study conducted by Fukudo et al. [6].

(4) Model details were not available (*n* = 6)

- The study carried out by Wählby et al. [7], Zahra Nasiri-Toosi et al. [8], Badri et al.[9], Wallemacq et al.[10], missed the details of parameters and formulas.
- The study by Pieter et al. [11] and Choi et al. [12] were excluded since their data just focused on investigating the effects of sampling time on model parameters.

**References**

1. Moes DJ, van der Bent SA, Swen JJ, et al. Population pharmacokinetics and pharmacogenetics of once daily tacrolimus formulation in stable liver transplant recipients [J]. European Journal of Clinical Pharmacology, 2016, 72(2):163-174.
2. Fukatsu S, Yano I, Igarashi T, et al. Population pharmacokinetics of tacrolimus in adult recipients receiving living-donor liver transplantation [J]. European Journal of Clinical Pharmacology, 2001, 57(6-7):479-484.
3. Fukudo M, Yano I, Fukatsu S, et al. Forecasting of Blood Tacrolimus Concentrations Based on the Bayesian Method in Adult Patients Receiving Living-Donor Liver Transplantation [J]. Clinical Pharmacokinetics, 2003, 42(13):1161-1178.
4. Valdivieso N, Oteo I, Valdivieso A, et al. Tacrolimus dose individualization in "de novo" patients after 10 years of experience in liver transplantation: pharmacokinetic considerations and patient pathophysiology [J]. International Journal of Clinical Pharmacology & Therapeutics, 2013, 51(7):606-614.
5. Oteo I, Lukas J C, Leal N, et al. Tacrolimus pharmacokinetics in the early post-liver transplantation period and clinical applicability via Bayesian prediction [J]. European Journal of Clinical Pharmacology, 2013, 69(1):65-74.
6. Fukudo M, Yano I, Yoshimura A, et al. Impact of MDR1 and CYP3A5 on the oral clearance of tacrolimus and tacrolimus-related renal dysfunction in adult living-donor liver transplant patients [J]. Pharmacogenetics and Genomics, 2008, 18(5):413-423.
7. Wählby U, Thomson A H, Milligan P A, et al. Models for time-varying covariates in population pharmacokinetic-pharmacodynamic analysis [J]. British Journal of Clinical Pharmacology, 2004, 58(4):367.
8. Nasiri-Toosi Z, Dashti-Khavidaki S, Nasiri-Toosi M, et al. Clinical Pharmacokinetics of Oral Versus Sublingual Administration of Tacrolimus in Adult Liver Transplant Recipients [J]. Experimental & Clinical Transplantation Official Journal of the Middle East Society for Organ Transplantation, 2012, 10:586.
9. Badri P S, Parikh A, Coakley E P, et al. Pharmacokinetics of Tacrolimus and Cyclosporine in Liver Transplant Recipients Receiving 3 Direct-Acting Antivirals as Treatment for Hepatitis C Infection [J]. Therapeutic Drug Monitoring, 2016, 38:640-645.
10. Wallemacq D P E, Verbeeck R K. Comparative Clinical Pharmacokinetics of Tacrolimus in Paediatric and Adult Patients [J]. Clinical Pharmacokinetics, 2001, 40:283-295.

11. Pieter L, Press R R, Den H J, et al. Flexible Limited Sampling Model for Monitoring Tacrolimus in Stable Patients Having Undergone Liver Transplantation with Samples 4 to 6 Hours After Dosing Is Superior to Trough Concentration [J]. Therapeutic Drug Monitoring, 2008, 30: 456-461.
12. Choi, Leena, Crainiceanu, et al. Practical recommendations for population PK studies with sampling time errors [J]. European Journal of Clinical Pharmacology 2013; 69:2055-2064.

*Appendix S3: Summarized information of all the identified studies*

Among all the identified studies, 62.5 % (n = 10) were reported in East Asian (six in China [8, 10, 12–15], two in Korea [16, 26], one each in Singapore [7] and Japan [1], respectively), and only 18.8% (n = 3) in Europe (two in France [4, 9], one in Spain [11]) and 18.8% (n = 3) in Australia [2, 3, 5] (Figure S1A).

Most of the included studies (87.5%) were single centre studies [1, 4–16] while only 12.5% were multicentre studies [2, 3]. Moreover, 43.8% (n = 7) of the studies had a small sample size of less than 50 patients [1, 3, 4, 6, 7, 9, 12], 37.5% (n = 6) of the studies used data from patients between 50 and 100 subjects [2, 5, 8, 11, 14, 16], and only 18.8% (n = 3) of the studies involved more than 100 participants [10, 13, 15] (Figure S1B).

Three kinds of bioassays were employed in previous studies, including MEIA in eleven studies [1, 4–6, 8, 10–14, 16], LC-MS in three studies [2, 3, 7], EMIT in one study [9], and LC-MS combined with MEIA in one study [15] (Figure S1C).

Additionally, the post-transplant period of follow-up varied greatly from one day to 2115 days after the liver transplantation (Figure S1D). Over half of the studies (62.5%, n = 10) were conducted during the first 6 months post-transplantation [1, 4, 5, 8, 9, 11–13, 15, 16], two of those were performed within 2 weeks post-transplantation [11, 16]. Others were conducted over six months post-transplantation. Only one study did not report the postoperative period [7].

*Supplementary table*

**Table S1 Results of prediction-based diagnostics**

| Models | MDPE (%) | MAPE (%) | $F_{20}$ (%) | $F_{30}$ (%) |
|---|---|---|---|---|
| *Published Studies* | | | | |
| Fukudo et al.(2003)[1] | -32.31 | 40.74 | 22.55 | 36.01 |
| Staatz et al.(2003)[2] | -28.51 | 37.50 | 27.45 | 39.16 |
| Dansirikul et al.(2004)[3] | -43.60 | 47.84 | 19.93 | 30.24 |
| Antignac et al. [4] | -43.58 | 46.57 | 18.18 | 29.20 |
| Zahir et al.(2005)[5] | -29.09 | 39.12 | 23.43 | 36.89 |
| Lee et al.(2006)[6] | 9.92 | 46.97 | 19.93 | 30.94 |
| Sam et al.(2006)[7] | 20.68 | 50.98 | 20.80 | 31.83 |
| Li et al.(2007)[8] | -11.56 | 43.43 | 27.10 | 37.41 |
| Benoit et al.(2008)[9] | 88.12 | 88.12 | 13.11 | 17.31 |
| Zhang et al.(2012)[10] | 4.17 | 30.96 | 35.31 | 49.13 |
| Oteo et al.(2013)[11] | -8.11 | 36.47 | 27.45 | 41.26 |
| Zhu et al.(2014)[12] | -33.88 | 38.81 | 22.38 | 36.54 |
| Lu et al.(2015)[13] | -19.53 | 37.21 | 28.50 | 39.86 |
| Zhu et al.(2015)[14] | -69.79 | 69.81 | 2.80 | 5.24 |
| Chen et al.(2017)[15] | -54.47 | 55.67 | 11.71 | 18.71 |
| Ji et al.(2018)[16] | -5.55 | 43.07 | 23.25 | 34.79 |
| *Impact of Model Structure* | | | | |
| MM (Base Model) | 4.42 | 34.34 | 29.72 | 43.01 |
| 2-CMT (Base Model) | -5.16 | 37.57 | 25.87 | 40.21 |
| 1-CMT (Base Model) | -1.35 | 38.89 | 25.87 | 38.99 |

MDPE (%), median prediction error; MAPE (%), median absolute prediction error; $F_{20}$ (%) and $F_{30}$ (%), the percentage of absolute prediction error within 20% and 30%, respectively; 1-CMT, one-compartmental model; 2-CMT, two-compartmental model; MM, Michaelis-Menten model.

**Table S2 Statistic test results of normalized prediction distribution error (NPDE) diagnostics**

| Study (publication year) | Mean (SE) | Variance (SE) | Skewness | Kurtosis | Wilcoxon signed rank test[a] | Fisher test[a] | Shapiro-Wilks test[a] | Global test[a] |
|---|---|---|---|---|---|---|---|---|
| *Published models* | | | | | | | | |
| Fukudo et al.(2003)[1] | 0.271 (0.040) | 0.921 (0.054) | 0.375 | 0.379 | 0.000*** | 0.176 | 0.000*** | 0.000*** |
| Staatz et al.(2003)[2] | 0.254 (0.025) | 0.369 (0.022) | 0.717 | 1.475 | 0.000*** | 0.000*** | 0.000*** | 0.000*** |
| Dansirikul et al.(2004)[3] | -1.416 (0.110) | 7.140 (0.420) | 0.920 | -0.811 | 0.000*** | 0.000*** | 0.000*** | 0.000*** |
| Antignac et al.(2005)[4] | 0.471 (0.030) | 0.531 (0.031) | 0.623 | 0.759 | 0.000*** | 0.000*** | 0.000*** | 0.000*** |
| Zahir et al.(2005)[5] | -0.218 (0.083) | 3.925 (0.230) | 0.105 | -0.873 | 0.008** | 0.000*** | 0.000*** | 0.000*** |
| Lee et al.(2006)[6] | -0.054 (0.036) | 0.761 (0.045) | 0.511 | 0.921 | 0.138 | 0.000*** | 0.000*** | 0.000*** |
| Sam et al.(2006)[7] | -0.390 (0.041) | 0.908 (0.055) | 0.168 | -0.396 | 0.000*** | 0.120 | 0.016* | 0.000*** |
| Li et al.(2007)[8] | 0.197 (0.041) | 0.949 (0.056) | 0.461 | 0.327 | 0.000*** | 0.392 | 0.000*** | 0.000*** |
| Benoit et al.(2008)[9] | -1.417 (0.082) | 3.819 (0.230) | 0.993 | 0.187 | 0.000*** | 0.000*** | 0.000*** | 0.000*** |
| Zhang et al.(2012)[10] | 0.079 (0.045) | 1.172 (0.069) | 0.434 | 0.327 | 0.082 | 0.005** | 0.000*** | 0.000*** |
| Oteo et al.(2013)[11] | -0.215 (0.066) | 2.457 (0.150) | 0.200 | -0.172 | 0.001** | 0.000*** | 0. 000*** | 0.000*** |
| Zhu et al.(2014)[12] | 0.394 (0.077) | 3.394 (0.200) | -0.160 | -0.745 | 0.000*** | 0.000*** | 0.000*** | 0.000*** |
| Lu et al.(2015)[13] | -0.152 (0.057) | 1.872 (0.110) | 0.085 | -0.290 | 0.008** | 0.000*** | 0.113 | 0.000*** |
| Zhu et al.(2015)[14] | 0.854 (0.075) | 3.175 (0.190) | -0.417 | -0.608 | 0.000*** | 0.000*** | 0.000*** | 0.000*** |
| Chen et al.(2017)[15] | -0.009 (0.091) | 4.718 (0.280) | 0.109 | -1.121 | 0.917 | 0.000*** | 0.000*** | 0.000*** |
| Ji et al.(2018)[16] | 0.136 (0.046) | 1.188 (0.070) | 0.603 | 0.416 | 0.003** | 0.003** | 0.000*** | 0.000*** |

SE, standard error;

[a]Data are presented as *P* value

***$P < 0.001$, **$P < 0.01$, *$P < 0.05$

**Table S3 Estimated parameters of investigated structural models**

| Parameter | Base Model |
|---|---|
| *Michaelis-Menten Model* | |
| OFV | 795.0 |
| Vm (mg day$^{-1}$) | 5.86 (12%) |
| Km (ng ml$^{-1}$) | 5.72 (23%) |
| BSV_Vm (CV%) | 15.6 (64%) |
| BSV_Km (CV%) | 66.7 (18%) |
| RUV | |
|     Proportional error (%) | 28.7 (15%) |
|     Additive error (ng ml$^{-1}$) | 0.569 (28%) |
| *One-Compartment Model* | |
| OFV | 1862.5 |
| CL/F (l h$^{-1}$) | 14.4 (6%) |
| V/F (l) | 258 (20%) |
| Ka (h$^{-1}$) | 4.48 (Fixed) |
| BSV_CL/F (CV%) | 31.6 (13%) |
| RUV | |
|     Proportional error (%) | 22.8 (24%) |
|     Additive error (ng ml$^{-1}$) | 2.19 (14%) |
| *Two-Compartment Model* | |
| OFV | 1909.4 |
| CL/F (l h$^{-1}$) | 13.4 (5%) |
| Vc/F (l) | 179 (8%) |
| Q/F (l h$^{-1}$) | 15 (Fixed) |
| Vp/F (l) | 300 (Fixed) |
| Ka (h$^{-1}$) | 4.48 (Fixed) |
| BSV_CL/F (CV%) | 30.7 (12%) |
| RUV | |
|     Proportional error (%) | 14.9 (68%) |
|     Additive error (ng ml$^{-1}$) | 2.64 (10%) |

BSV, between subject variability; CL/F, apparent clearance; CV, coefficient of variation; Ka, absorption rate constant; Km, Michales-Menten constant equal to the steady-state trough concentration at half-maximum dose rate; OFV, objective function value; Q/F, apparent inter-compartmental clearance; RUV, residual unexplained variability; Vc/F, apparent central volume of distribution; Vm, the maximal dose rate (daily dose) at steady state; Vp/F, apparent peripheral volume of distribution.

Ka, Q/F, and Vp/F were fixed. All estimations are provided as model estimate (relative standard error, RSE%). The BSV on Km and Vm in Michales-Menten model or CL/F in other linear compartment models were described with exponential error model. The residual error in all the structural models was modelled using a mixed exponential-additive model.

**Table S4 Results of Bayesian forecasting**

| Models | MDIPE (%) | | | | | MAIPE (%) | | | | | IF$_{20}$ (%) | | | | | IF$_{30}$ (%) | | | | |
|---|---|---|---|---|---|---|---|---|---|---|---|---|---|---|---|---|---|---|---|---|
| | P$_0$ | P$_1$ | P$_2$ | P$_3$ | P$_4$ | P$_0$ | P$_1$ | P$_2$ | P$_3$ | P$_4$ | P$_0$ | P$_1$ | P$_2$ | P$_3$ | P$_4$ | P$_0$ | P$_1$ | P$_2$ | P$_3$ | P$_4$ |
| *Published studies* | | | | | | | | | | | | | | | | | | | | |
| Fukudo et al.(2003)[1] | -13 | 2 | 8 | 11 | 15 | 37 | 28 | 30 | 33 | 32 | 27 | 36 | 35 | 33 | 33 | 42 | 51 | 49 | 49 | 45 |
| Staatz et al.(2003)[2] | -20 | -12 | -8 | -4 | -1 | 41 | 34 | 35 | 30 | 30 | 25 | 26 | 32 | 37 | 38 | 35 | 40 | 42 | 50 | 50 |
| Dansirikul et al.(2004)[3] | -28 | 14 | 37 | 56 | 114 | 35 | 26 | 41 | 57 | 114 | 27 | 37 | 27 | 12 | 5 | 39 | 57 | 42 | 24 | 8 |
| Antignac et al.(2005)[4] | -44 | -29 | -21 | -21 | -20 | 45 | 32 | 27 | 26 | 26 | 21 | 27 | 38 | 36 | 36 | 30 | 42 | 56 | 55 | 56 |
| Zahir et al.(2005)[5] | -23 | 3 | 1 | 6 | 11 | 39 | 28 | 26 | 30 | 32 | 18 | 37 | 39 | 39 | 38 | 36 | 51 | 55 | 50 | 48 |
| Lee et al.(2006)[6] | 13 | 11 | 9 | 18 | 23 | 49 | 31 | 35 | 33 | 37 | 20 | 32 | 33 | 29 | 26 | 33 | 49 | 44 | 46 | 42 |
| Sam et al.(2006)[7] | 42 | 27 | 16 | 33 | 31 | 56 | 36 | 37 | 43 | 47 | 17 | 27 | 40 | 35 | 30 | 26 | 42 | 47 | 38 | 38 |
| Li et al.(2007)[8] | -6 | -4 | 3 | 3 | 9 | 42 | 34 | 33 | 34 | 36 | 30 | 32 | 36 | 33 | 31 | 39 | 45 | 48 | 40 | 40 |
| Benoit et al.(2008)[9] | 167 | 48 | 70 | 86 | 87 | 167 | 48 | 70 | 86 | 87 | 5 | 20 | 13 | 8 | 6 | 6 | 31 | 19 | 13 | 8 |
| Zhang et al.(2012)[10] | 20 | 8 | 12 | 11 | 14 | 28 | 26 | 25 | 23 | 28 | 37 | 42 | 42 | 45 | 37 | 51 | 54 | 60 | 58 | 51 |
| Oteo et al.(2013)[11] | 6 | 20 | 21 | 27 | 28 | 40 | 28 | 30 | 37 | 39 | 25 | 30 | 33 | 35 | 36 | 36 | 54 | 50 | 45 | 43 |
| Zhu et al.(2014)[12] | -28 | -13 | -6 | -1 | 3 | 33 | 25 | 24 | 22 | 23 | 25 | 42 | 42 | 44 | 43 | 43 | 60 | 63 | 65 | 64 |
| Lu et al.(2015)[13] | -2 | 0 | 6 | 10 | 15 | 37 | 30 | 29 | 30 | 31 | 26 | 32 | 36 | 39 | 36 | 36 | 50 | 51 | 51 | 50 |
| Zhu et al.(2015)[14] | -70 | -23 | -14 | -10 | -7 | 70 | 25 | 23 | 22 | 23 | 4 | 38 | 45 | 45 | 46 | 7 | 60 | 67 | 63 | 61 |
| Chen et al.(2017)[15] | -49 | 5 | 15 | 25 | 34 | 50 | 25 | 32 | 34 | 40 | 21 | 31 | 35 | 30 | 29 | 25 | 60 | 46 | 44 | 37 |
| Ji et al.(2018)[16] | 4 | 8 | 14 | 18 | 25 | 43 | 36 | 37 | 35 | 38 | 26 | 33 | 32 | 33 | 31 | 35 | 45 | 43 | 43 | 42 |
| *Impact of Model Strucutre* | | | | | | | | | | | | | | | | | | | | |
| MM (Base Model) | -12 | -12 | -11 | -12 | -15 | 35 | 21 | 23 | 25 | 26 | 29 | 49 | 40 | 40 | 35 | 37 | 74 | 67 | 65 | 57 |
| 2-CMT (Base Model) | 17 | 8 | 11 | 16 | 19 | 40 | 32 | 34 | 33 | 37 | 24 | 29 | 37 | 36 | 33 | 38 | 48 | 46 | 46 | 43 |
| 1-CMT (Base Model) | 15 | 5 | 7 | 7 | 10 | 44 | 32 | 33 | 32 | 35 | 23 | 31 | 37 | 38 | 35 | 32 | 46 | 49 | 49 | 48 |

MDIPE (%), median individual prediction error; MAIPE (%), median individual absolute prediction error; IF$_{20}$ (%) and IF$_{30}$ (%), the percentage of individual prediction error within ± 20% and ± 30%, respectively. 1-CMT, one-compartment model; 2-CMT, two-compartment model; MM, Michaelis-Menten model; n, numbers of priors; Pn, predictions with no prior observation ($n = 0$) and with 1–4 prior observations ($n = 1–4$), respectively.

**References**

1. Fukudo M, Yano I, Fukatsu S, Saito H, Uemoto S, Kiuchi T, Tanaka K, Inui K. Forecasting of blood tacrolimus concentrations based on the Bayesian method in adult patients receiving living-donor liver transplantation[J]. Clin Pharmacokinet 2003; 42: 1161-78.
2. Staatz CE, Willis C, Taylor PJ, Lynch SV, Tett SE. Toward better outcomes with tacrolimus therapy: population pharmacokinetics and individualized dosage prediction in adult liver transplantation[J]. Liver Transpl 2003; 9: 130-7.
3. Dansirikul C, Staatz CE, Duffull SB, Taylor PJ, Lynch SV, Tett SE. Sampling times for monitoring tacrolimus in stable adult liver transplant recipients[J]. Ther Drug Monit 2004; 26: 593-9.
4. Antignac M, Hulot JS, Boleslawski E, Hannoun L, Touitou Y, Farinotti R, Lechat P, Urien S. Population pharmacokinetics of tacrolimus in full liver transplant patients: modelling of the post-operative clearance[J]. Eur J Clin Pharmacol 2005; 61: 409-16.
5. Zahir H, McLachlan AJ, Nelson A, McCaughan G, Gleeson M, Akhlaghi F. Population pharmacokinetic estimation of tacrolimus apparent clearance in adult liver transplant recipients[J]. Ther Drug Monit 2005; 27: 422-30.
6. Lee JY, Hahn HJ, Son IJ, Suh KS, Yi NJ, Oh JM, Shin WG. Factors affecting the apparent clearance of tacrolimus in Korean adult liver transplant recipients[J]. Pharmacotherapy 2006; 26: 1069-77.

7. Sam WJ, Tham LS, Holmes MJ, Aw M, Quak SH, Lee KH, Lim SG, Prabhakaran K, Chan SY, Ho PC. Population pharmacokinetics of tacrolimus in whole blood and plasma in asian liver transplant patients[J]. Clin Pharmacokinet 2006; 45: 59-75.

8. Li D, Lu W, Zhu JY, Gao J, Lou YQ, Zhang GL. Population pharmacokinetics of tacrolimus and CYP3A5, MDR1 and IL-10 polymorphisms in adult liver transplant patients[J]. J Clin Pharm Ther 2007; 32: 505-15.

9. Blanchet B, Duvoux C, Costentin CE, Barrault C, Ghaleh B, Salvat A, Jouault H, Astier A, Tod M, Hulin A. Pharmacokinetic-pharmacodynamic assessment of tacrolimus in liver-transplant recipients during the early post-transplantation period[J]. Ther Drug Monit 2008; 30: 412-8.

10. Zhang XQ, Wang ZW, Fan JW, Li YP, Jiao Z, Gao JW, Peng ZH, Liu GL. The impact of sulfonylureas on tacrolimus apparent clearance revealed by a population pharmacokinetics analysis in Chinese adult liver-transplant patients[J]. Ther Drug Monit 2012; 34: 126-33.

11. Oteo I, Lukas JC, Leal N, Suarez E, Valdivieso A, Gastaca M, Ortiz de Urbina J, Calvo R. Tacrolimus pharmacokinetics in the early post-liver transplantation period and clinical applicability via Bayesian prediction[J]. Eur J Clin Pharmacol 2013; 69: 65-74.

12. Zhu L, Wang H, Sun X, Rao W, Qu W, Zhang Y, Sun L. The Population Pharmacokinetic Models of Tacrolimus in Chinese Adult Liver Transplantation Patients[J]. J Pharm (Cairo) 2014; 2014: 713650.

13. Lu YX, Su QH, Wu KH, Ren YP, Li L, Zhou TY, Lu W. A population pharmacokinetic study of tacrolimus in healthy Chinese volunteers and liver transplant patients[J]. Acta Pharmacol Sin 2015; 36: 281-8.

14. Zhu L, Yang J, Zhang Y, Jing Y, Zhang Y, Li G. Effects of CYP3A5 genotypes, ABCB1 C3435T and G2677T/A polymorphism on pharmacokinetics of Tacrolimus in Chinese adult liver transplant patients[J]. Xenobiotica 2015; 45: 840-6.

15. Chen B, Shi HQ, Liu XX, Zhang WX, Lu JQ, Xu BM, Chen H. Population pharmacokinetics and Bayesian estimation of tacrolimus exposure in Chinese liver transplant patients[J]. J Clin Pharm Ther 2017; 42: 679-88.

16. Ji E, Kim MG, Oh JM. CYP3A5 genotype-based model to predict tacrolimus dosage in the early postoperative period after living donor liver transplantation[J]. Ther Clin Risk Manag 2018; 14: 2119-26.

Figure S1

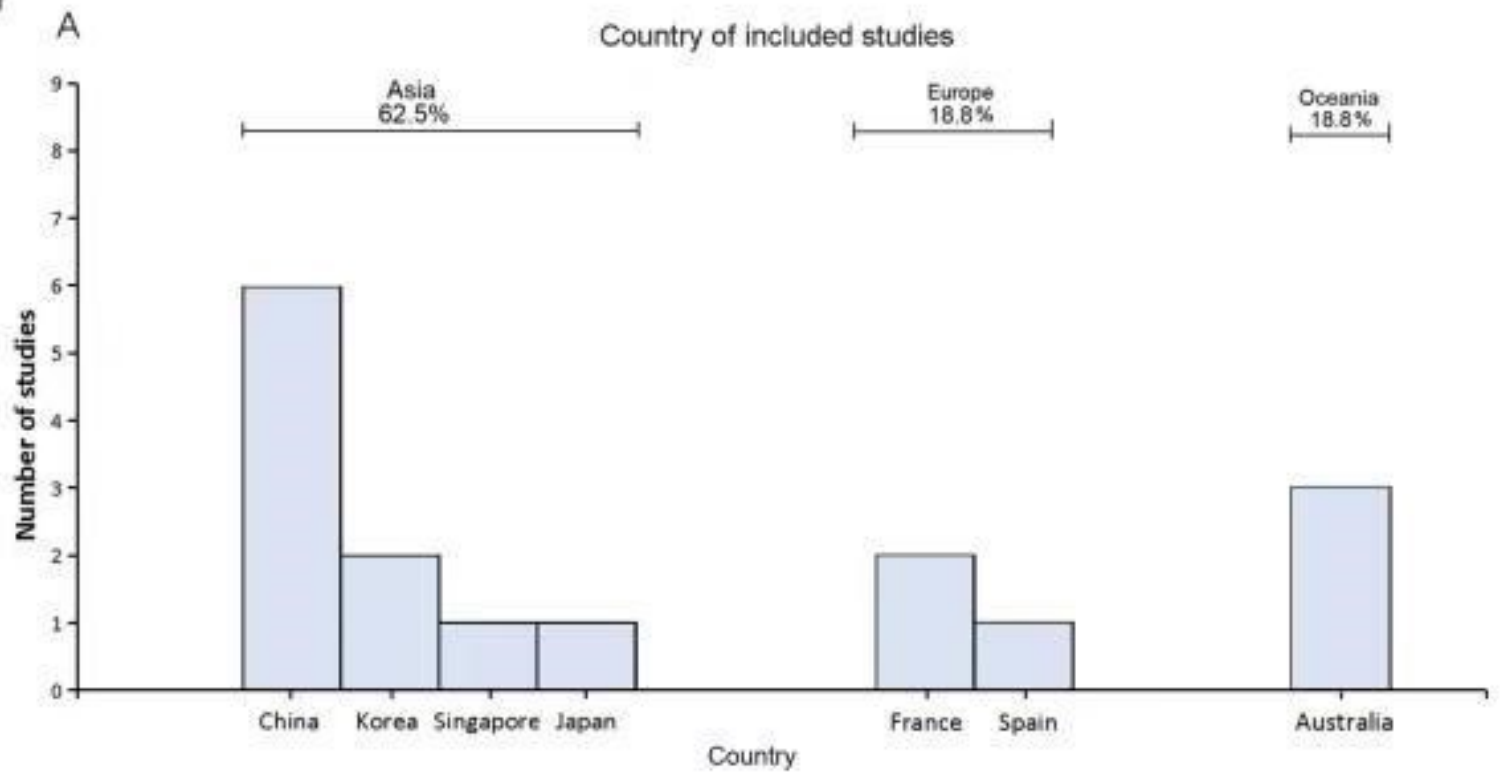
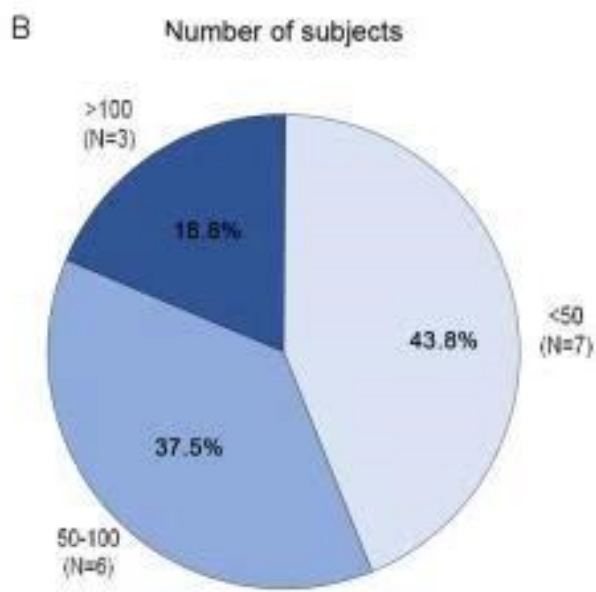
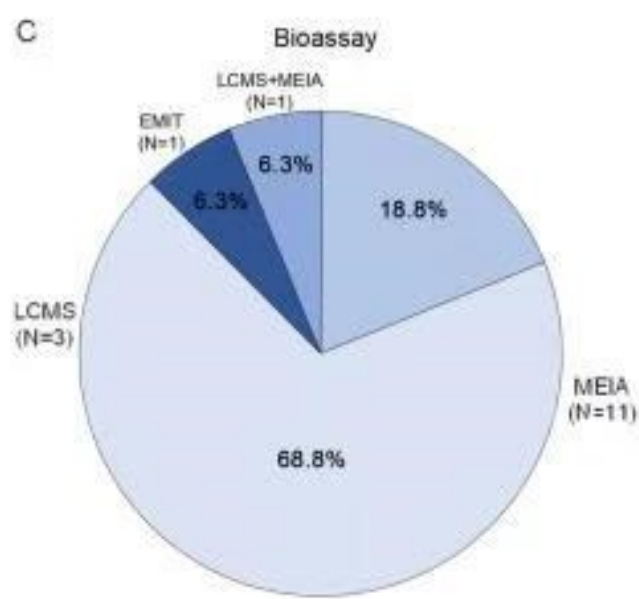
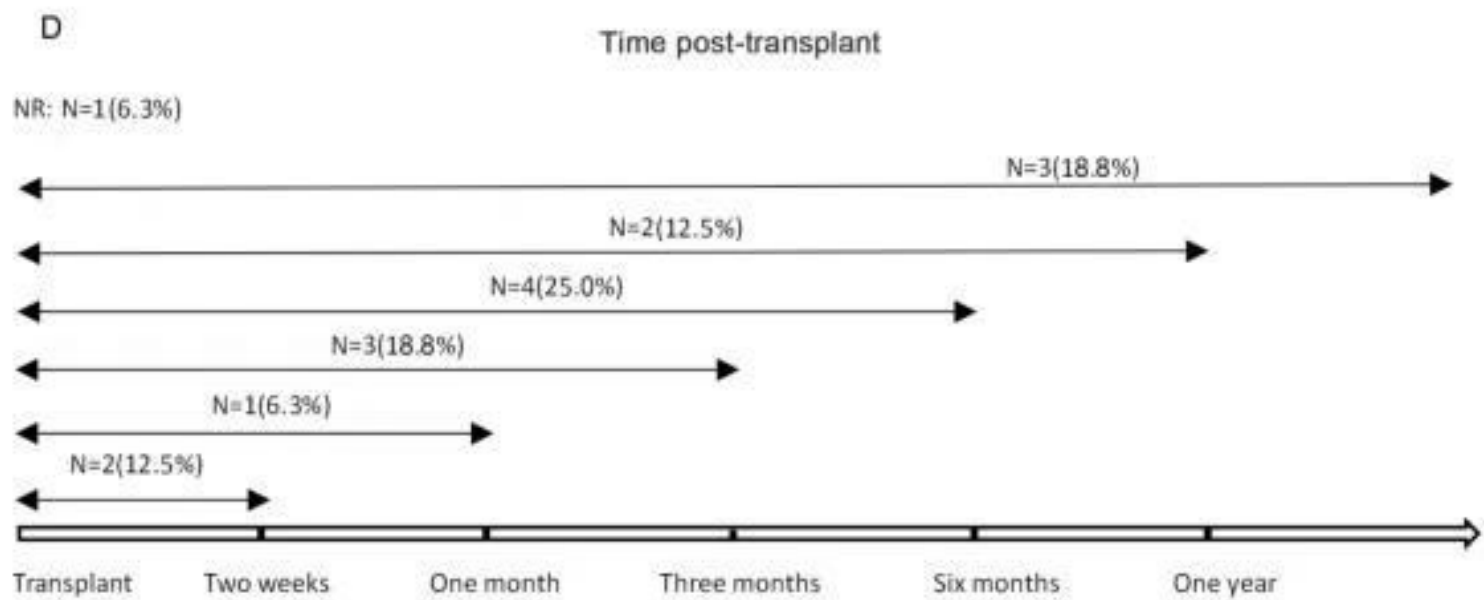

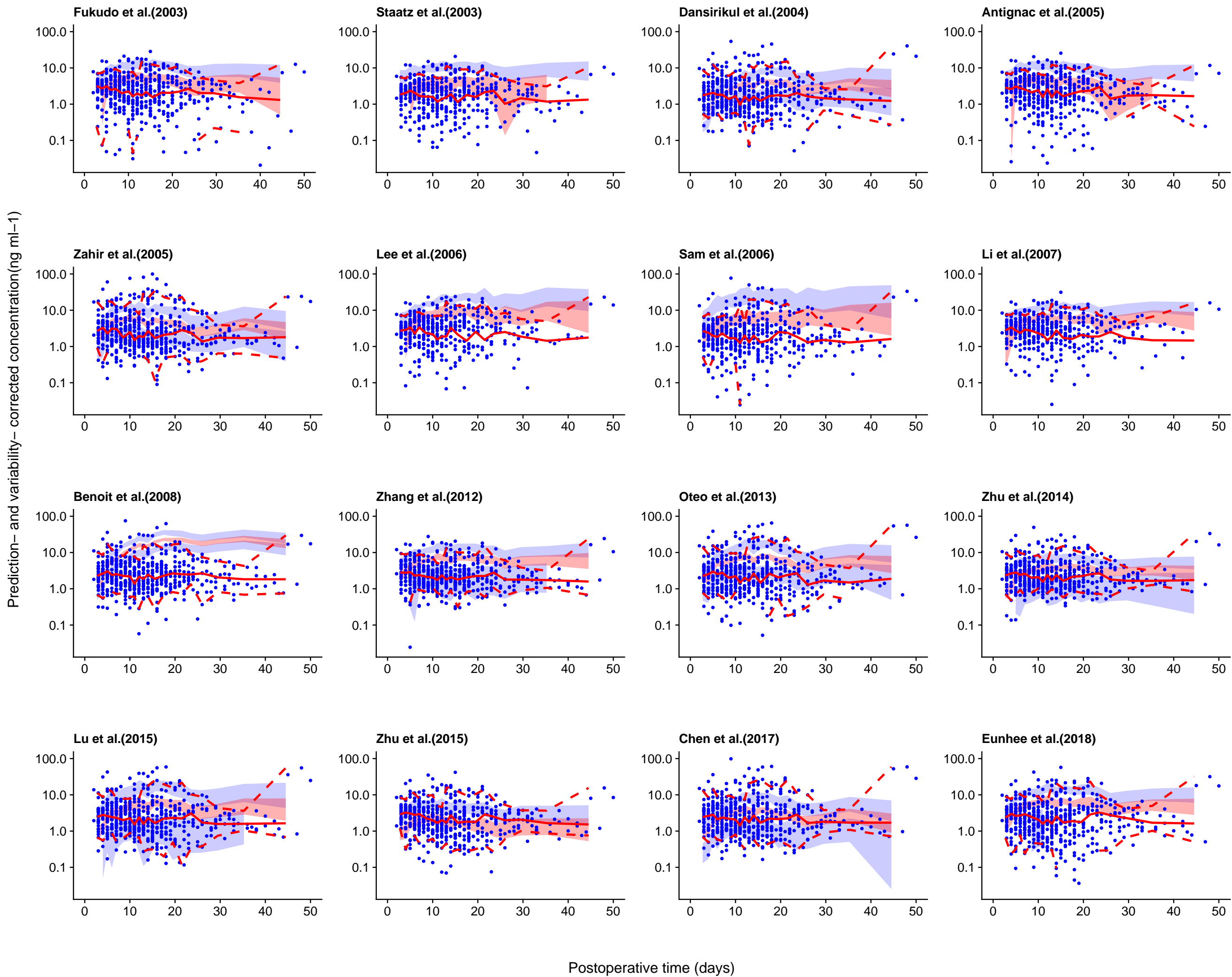

**Figure S2B**

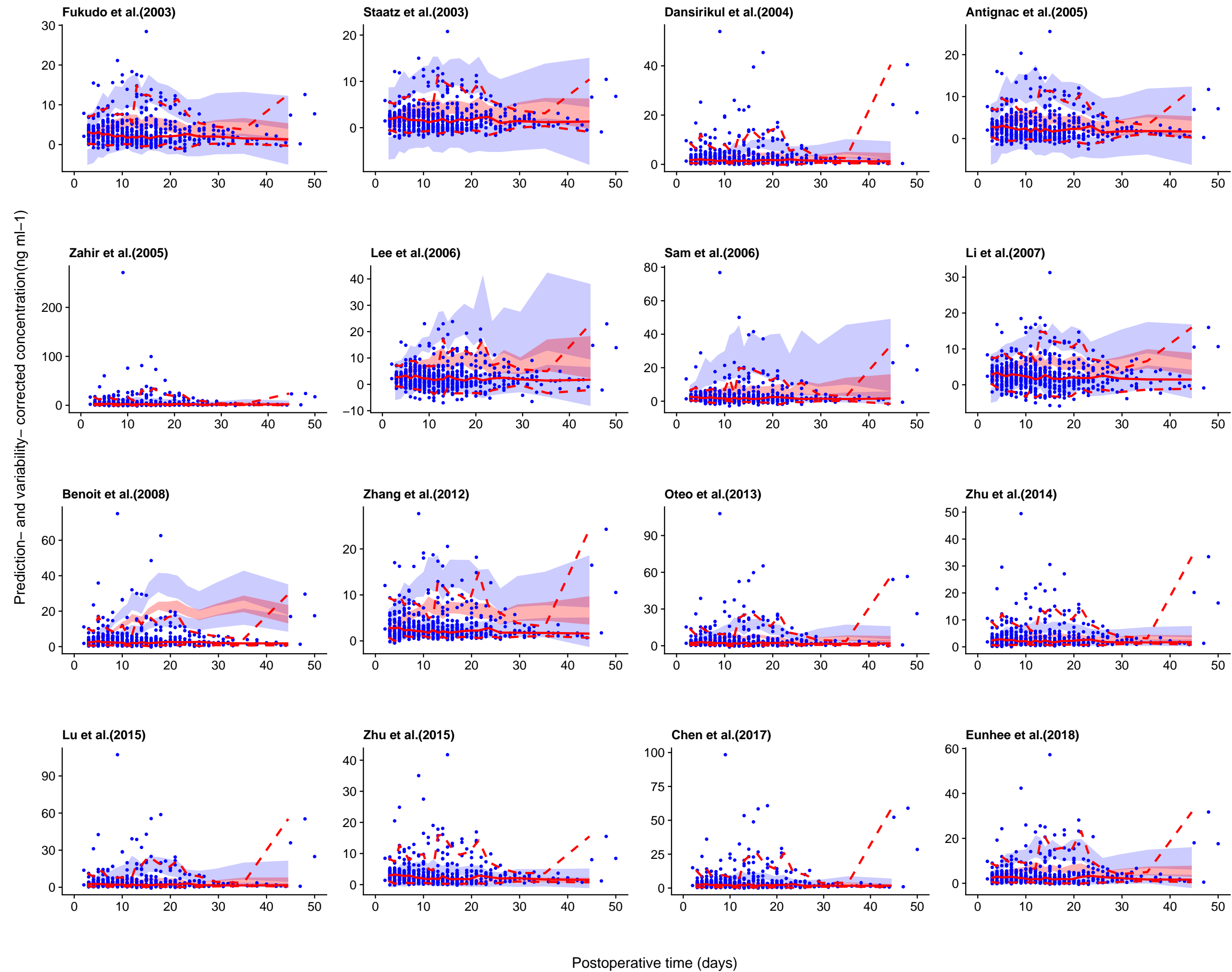

Figure S3A

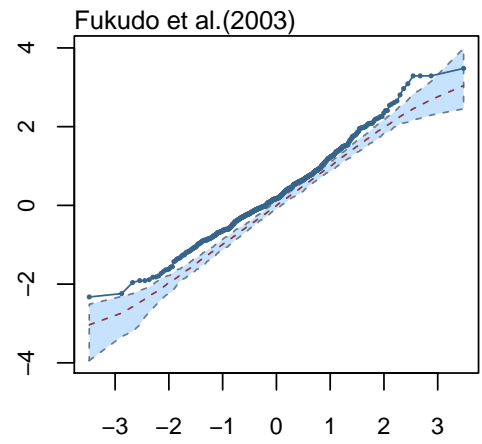 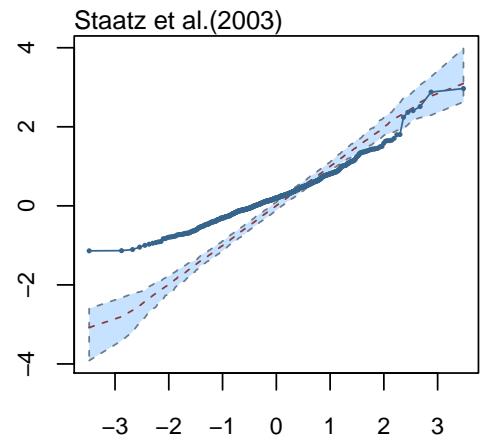 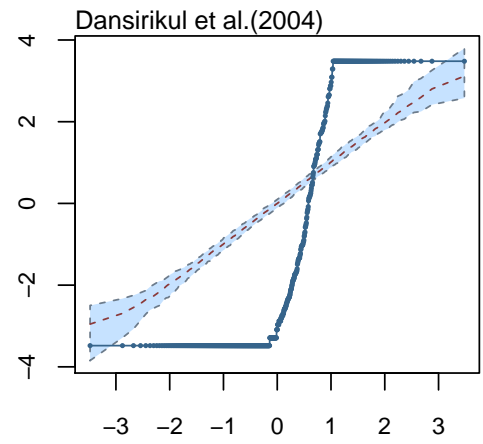 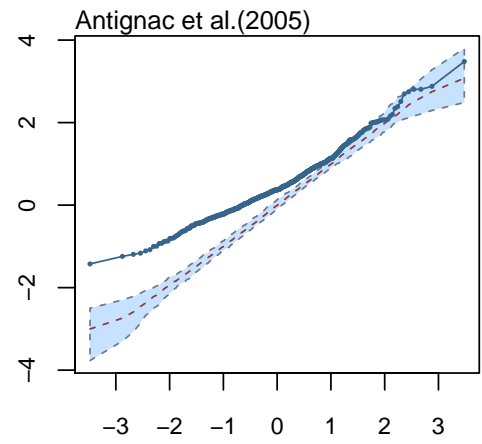
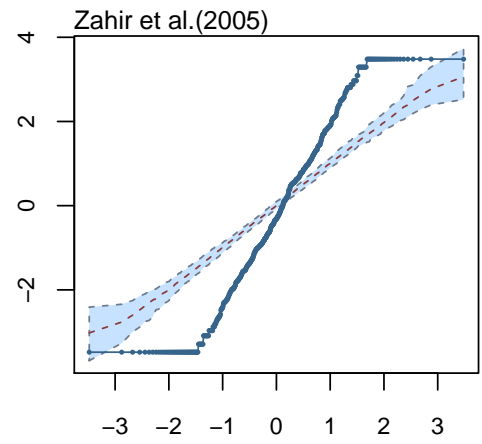 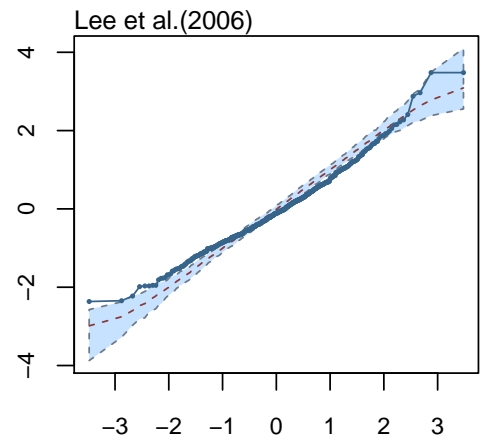 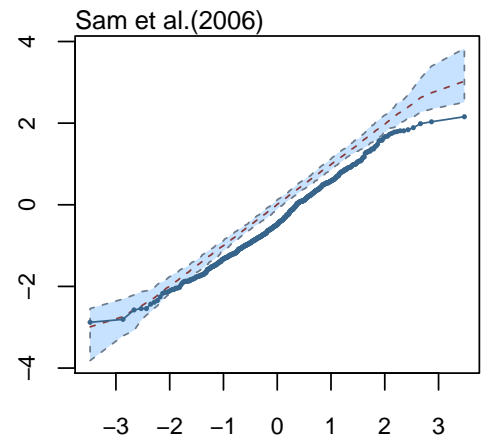 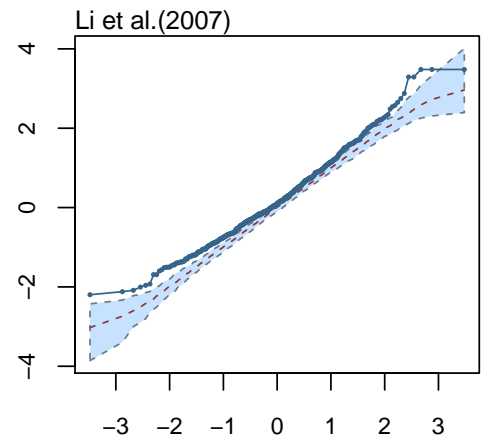
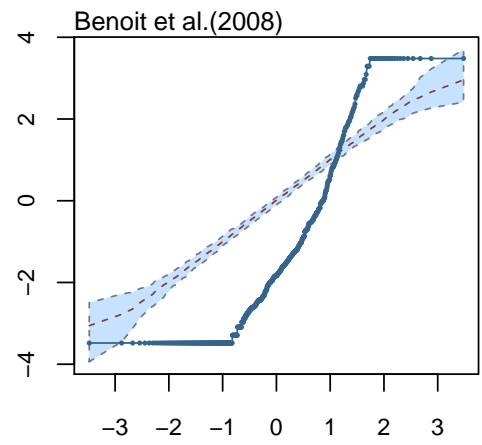 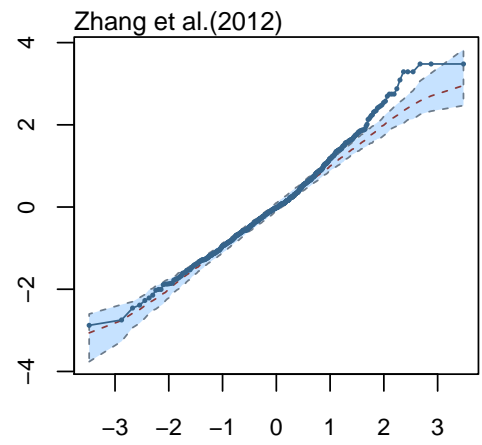 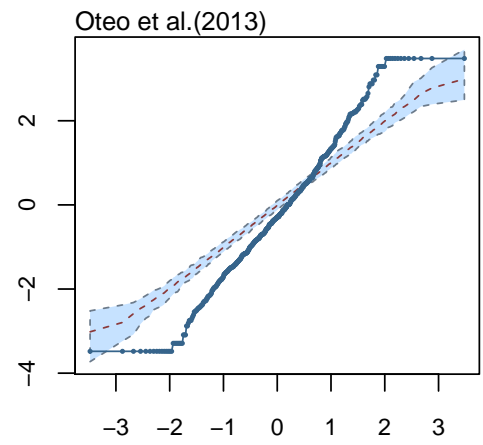 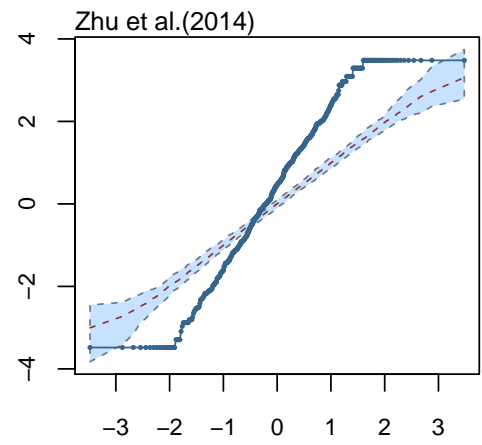
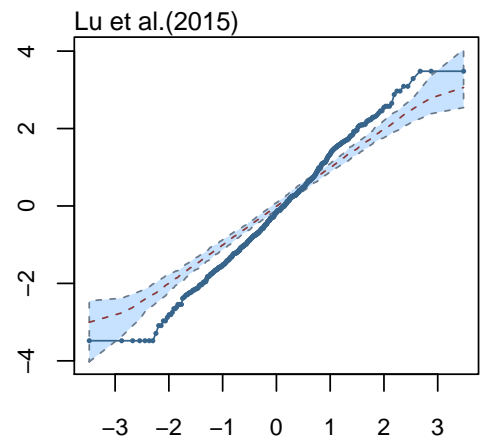 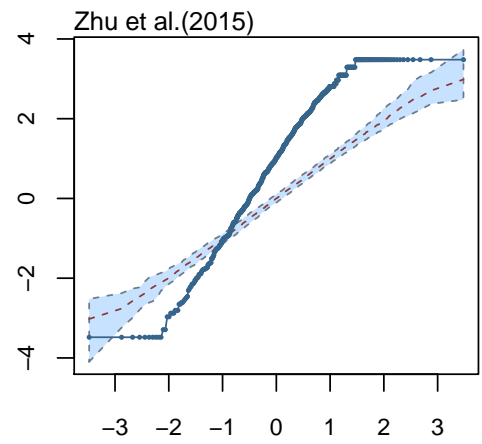 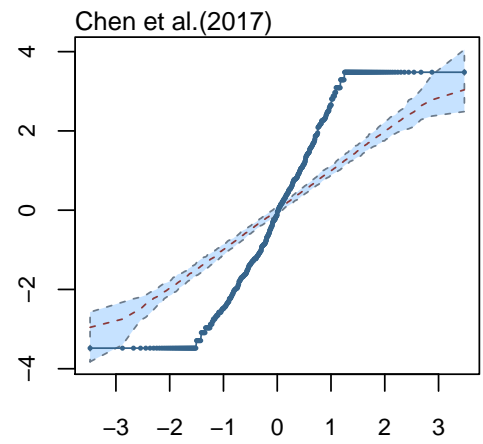 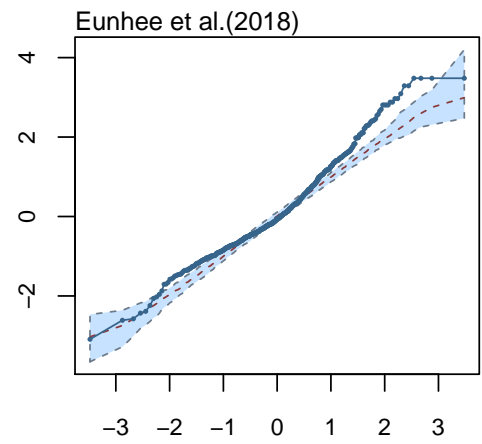

**Theoratical quantiles**

**Sample quantiles (npde)**

Figure S3B

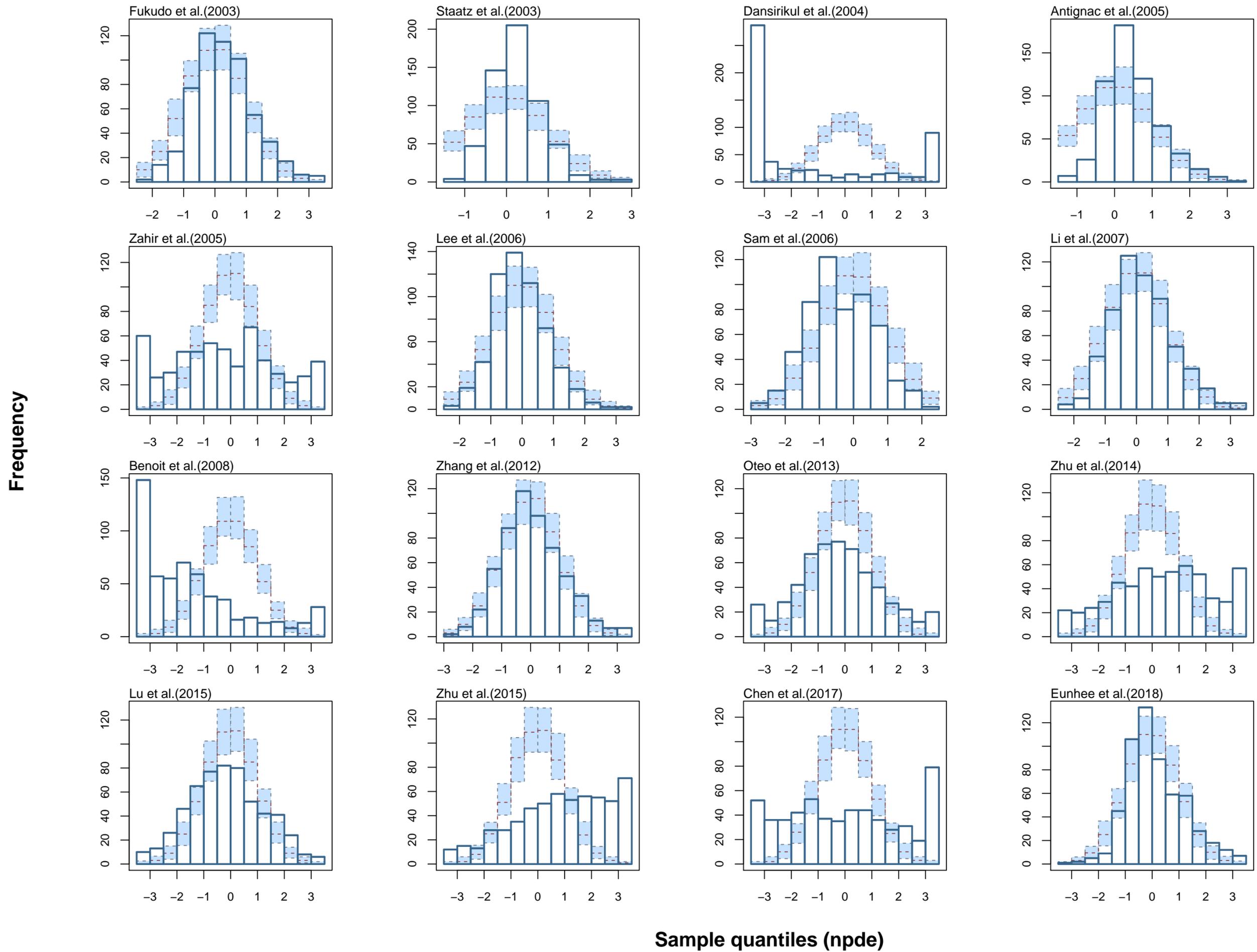

Figure S3C

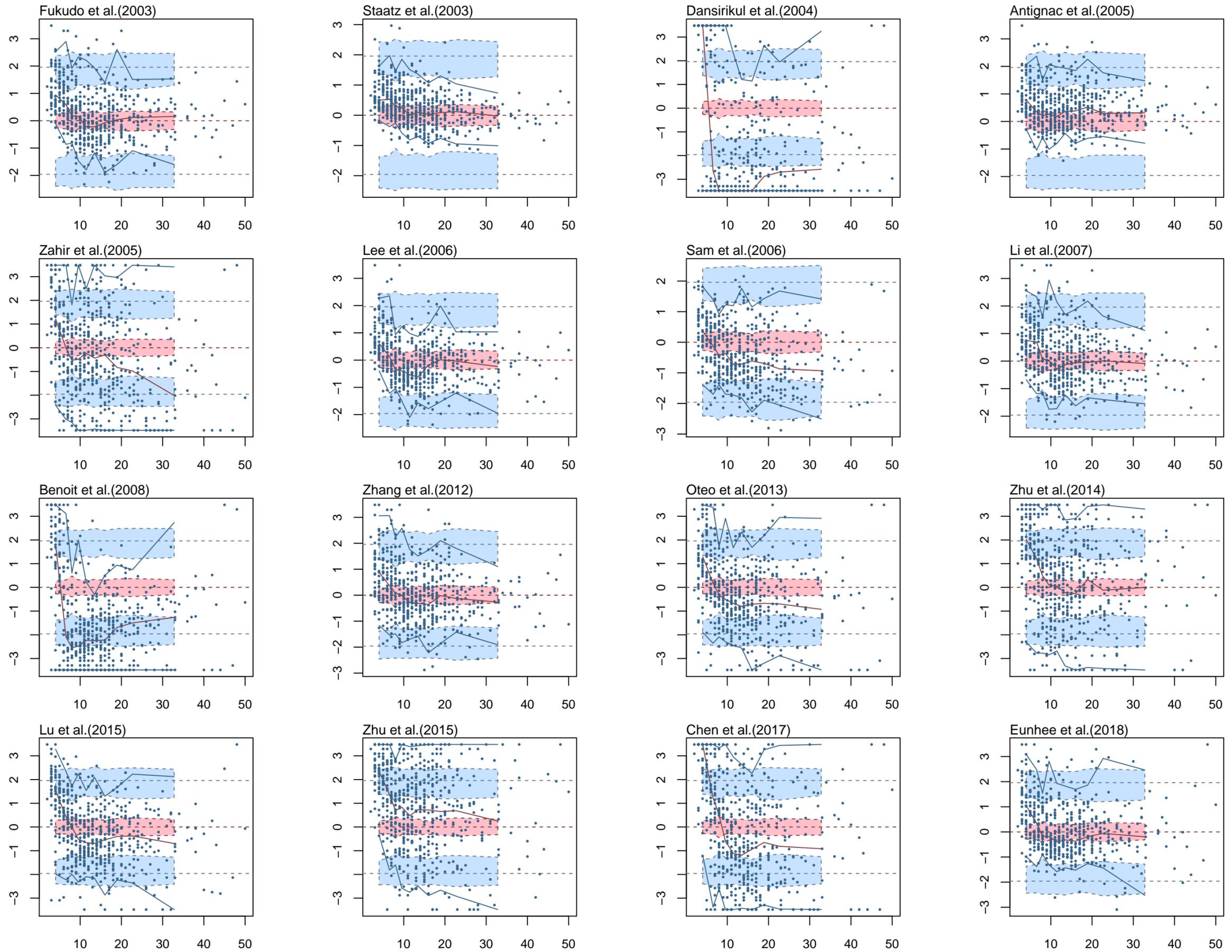

# Figure S3D

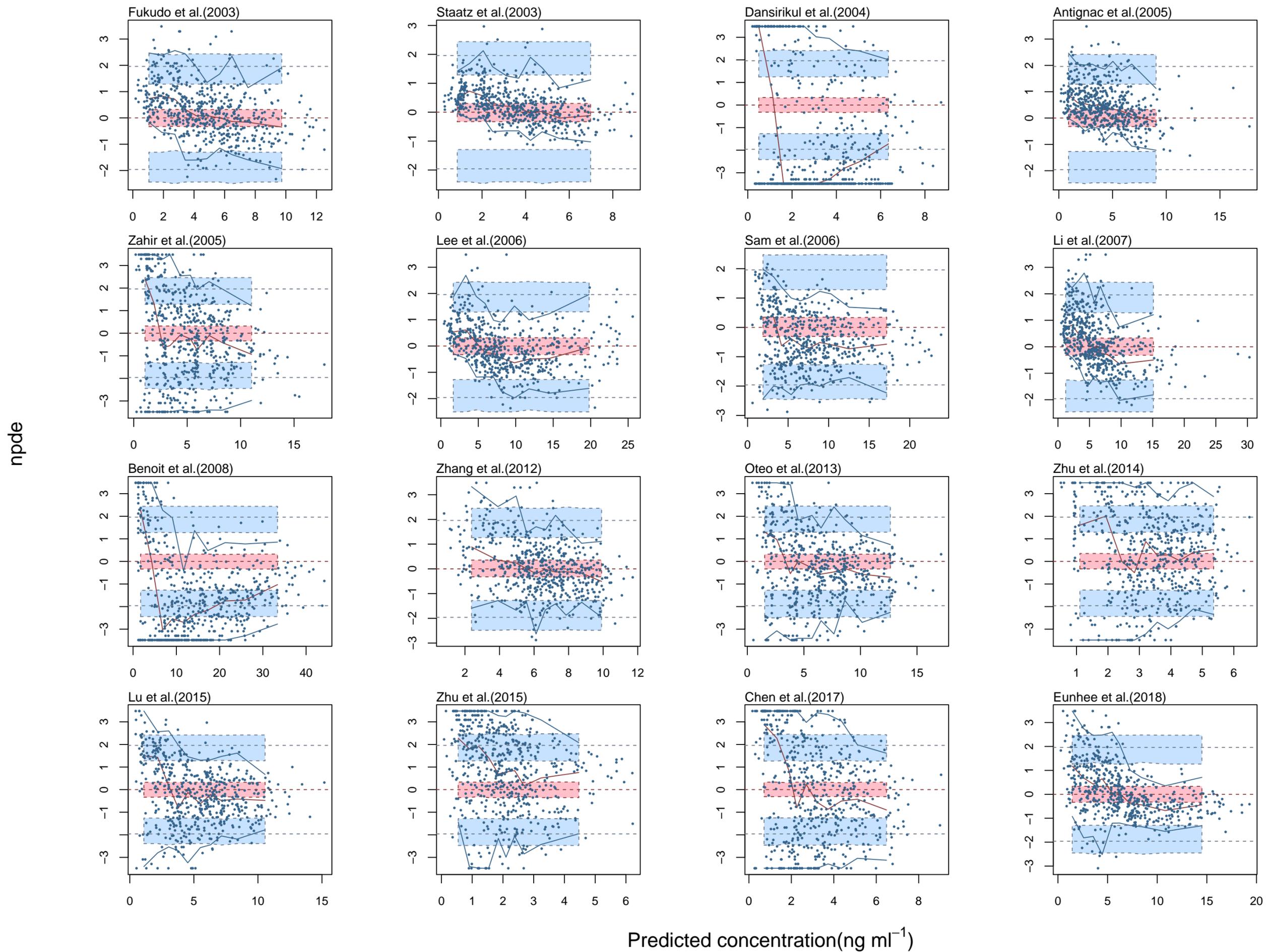